%% file: main.tex
\documentclass[bibyear]{aa}  

\usepackage{stfloats}
\usepackage{graphicx}
\usepackage{float}
\usepackage[colorlinks=true, allcolors=blue]{hyperref}
\usepackage{natbib}
\bibpunct{(}{)}{;}{a}{}{,}
\usepackage{threeparttable}
\usepackage{tabularx}
\usepackage{lscape}
\usepackage[version=4]{mhchem}

\AtBeginDocument{}%
\AtBeginDocument{}%
\AtBeginDocument{}%
\AtBeginDocument{}%
\AtBeginDocument{}%

\usepackage{txfonts}
\usepackage{xcolor}

\begin{document} 

\title{Thermal processing of primordial pebbles in evolving protoplanetary disks}
\titlerunning{Thermal processing in protoplanetary disks}    

\author{Mar\'ia Jos\'e Colmenares
      \inst{1,}\thanks{Corresponding author} 
      \and Michiel Lambrechts \inst{2,}\inst{3}
      \and Elishevah van Kooten \inst{2}
     \and Anders Johansen \inst{2,}\inst{3}}

\institute{Department of Astronomy, University of Michigan, 323 West Hall, 1085 S. University Avenue, Ann Arbor, MI 48109, USA\\ 
\email{mjcolmen@umich.edu}
     \and
         Center for Star and Planet Formation and Natural History Museum of Denmark, Globe Institute, University of Copenhagen, Øster Voldgade 5–7, 1350 Copenhagen, Denmark
    \and 
    Lund Observatory, Department of Astronomy and Theoretical Physics, Lund University, Box 43, 22100 Lund, Sweden}
    
\date{Received August 22, 2023; accepted February 06, 2024}

\abstract{During protoplanetary disk formation, dust grains located in the outer disk retain their pristine icy composition, while solids in the inner stellar-heated disk undergo volatile loss. This process may have left a fossil record in Solar System material showing different nucelosynthetic imprints that have been attributed to different degrees of thermal processing. However, it remains unclear how a large mass fraction of thermally-processed inner-disk pebbles is produced and how these grains are subsequently transported throughout the disk. In this work we numerically investigate the evolution in time of a two-component pebble disk, consisting of pristine pebbles and those that underwent ice sublimation. We find that stellar outbursts exceeding 1000 times the solar luminosity are efficient in thermally altering, through ice sublimation, a large mass fraction of pebbles (around 80\%). After the establishment of this initial radial dust composition gradient throughout the disk, the subsequent mixing and inward drift of pristine outer-disk pebbles alter the inner disk bulk composition from processed to more unprocessed in time. Therefore,  if processed pebbles without ice mantles have an isotopic composition similar to ureilite meteorites from the inner Solar System, inner-disk minor bodies forming from the early pebble flux ($<$1Myr) will be isotopically ureilite-like, while later-formed bodies will be increasingly admixed with the signature of the late incoming CI chondrite-like unprocessed pebbles. This appears to be largely consistent with the trend seen between the accretion age of different meteoric classes and their different stable isotope composition anomalies (in µ\ce{^{54}Cr}, µ\ce{^{48}Ca}, µ\ce{^{30}Si}, µ\ce{^{58}Ni}), but further work may be needed to explain the role of isotopically anomalous refractory inclusions {and anomaly trends in other elements}. Our findings further supports early thermal-processing of ice mantles via stellar outbursts that are common around young sun-like stars.}

\keywords{protoplanetary disks -- meteorites, meteors, meteoroids -- planets and satellites: formation -- planets and satellites: composition -- minor planets, asteroids: general}

\maketitle

\section{Introduction}

Young stars are surrounded by protoplanetary disks that form as an outcome of the collapse of dense molecular cloud cores. Such disks are composed of a dominant hydrogen/helium gas component and a minor solid dust component which makes up only roughly one percent of the total mass fraction. These solid particles efficiently grow to mm-sizes, as observed in the protoplanetary disk survey of star-forming regions by the Atacama Large Millimeter Array (ALMA) \citep{ansdell2016,andrews2018,tychoniec2020}. The main mechanism that limits the size of these pebbles is believed to be collisional fragmentation, which occurs for mm-sized particles at characteristic collision velocities of 1-10 m/s \citep{blum2008,musiolik2019}. 
Pebbles in this size range also experience angular momentum loss due to gas drag, which causes them to spiral inward within disk lifetimes of a few Myr \citep{weidensch1977}. Indeed, disks are typically formed with hundreds of Earth masses in pebbles, but this mass budget decreases to less than an Earth mass after several Myr, consistent with fragmentation-limited pebble drift \citep{brauer2008,birnstiel2010,Appelgren2023}.

It its not well understood how large-scale pebble drift affected the composition of minor bodies and rocky planets in the Solar System. Measurements of the isotopic signature of meteorites show a clear distinction between the so-called carbonaceous (CC) and non-carbonaceous (NC) meteorite groups. This CC group is often related to more volatile-rich outer-disk parent bodies similar to the pristine solar-like CI-chondrites. It carries a neutron-rich isotopic signature for elements such as Cr, Ca, Ti, and Si \citep{trinquier2007,warren2011,schiller2018,kruijer2020, onyett2023}. In contrast, the NC group is assumed to sample more volatile-depleted bodies originating from the inner Solar System. This group goes paired with an opposing isotopic imprint, which is neutron poor for these elements. However, {groups like} ordinary chondrites (OC), typically classified as NC, are close to intermediate in the NC-CC isotope space and therefore it may be more accurate to speak of a gradient than a true isotopic dichotomy \citep[][see also \autoref{implications}]{onyett2023}. Moreover, there is a third isotopically distinct group consisting of early-formed highly refractory inclusions in meteorites, the so-called Ca-Al-rich inclusions (CAIs) and amoeboid olivine aggregates (AOAs). 
Intriguingly, their isotopic composition is far more neutron enriched compared to the CC group, including Ivuna-type carbonaceous chondrites (CI), in elements like Cr and Ti \citep{trinquier2009}. 
It is currently not clear if this implies that (i) the CC group should be seen as an isotopic mixture of NC and precursor material of CAIs and AOAs \citep{nanne2019}, or (ii) that the NC group should be seen as processed material that lost a neutron-rich component to CAI\slash AOA-like material \citep{trinquier2009}, or (iii) that  neither of these interpretations are fully correct. In any case, it does appear that (fragments of) these refractory inclusions, or material isotopically similar to it, were radially well spread and included in both NC and CC groups, including even in CI-like parent bodies like Ryugu \citep{Kawasaki_2022}.
More in-depth reviews on the nucleosynthetic isotope compositions of meteorites can be found in \citet{qin2016}, \citet{Dauphas2016AREPS}, \citet{Kleine2020} and \citet{kruijer2020}.

{The existence of bodies with such distinct isotopic compositions has been interpreted to indicate either a \emph{spatial separation}: an initial radial isotopic gradient that was maintained during disk evolution by somehow halting pebble drift
or a \emph{temporal evolution}: a change in time of the composition of solids available for planet formation.
This latter view appears to be supported by the evolving isotope composition in the terrestrial region, as traced by the refractory elements 
\ce{^{48}Ca} and \ce{^{30}Si} \citep{schiller2018,onyett2023}.
Alternatively, it has been proposed that the presence of Jupiter spatially separated inner and outer disk reservoirs for a couple of million years \citep{vankooten2016,kruijer2017,alibert2018}, which may be challenging given the efficient transport of pebble fragments across planetary gaps \citep{stammler2023,Kalyaan2023}.
}

{
These different views are linked to how we understand the formation of the Earth and the other terrestrial planets, because their isotopic composition is the cumulative product of the accreted pebbles and planetesimals in terrestrial-planet forming zone \citep{Lichtenberg2021,johansen2021}. A substantial contribution to Earth of late-accreted CC-like material with outer disk origins would favor the temporal evolution case.
Studies tracing the refractory elements \ce{^{48}Ca} and \ce{^{30}Si}, major planetary components, allow for a substantial 30\% CC mass fraction contribution to the Earth and approximately 10\% for Mars \citep{schiller2018,onyett2023}, broadly in agreement with earlier work based on Cr and Ti isotopic anomalies \citep{warren2011}. 
Recent works exploring isotope anomalies fore more volatile elements, K \citep{Nie2023} and Zn \citep{Steller2022,Savage2022,Martins2023}, generally argue for a lower CC contribution for the Earth, around $10$\,\% .
When considering s-process elements Zr and Mo, \cite{burkhardt2021} argue that the Earth may have formed shielded from outer disk material and instead invoke a contribution from now lost, and thus unsampled, inner-disk material. However, as argued in \citet{onyett2023}, Mo and other elements with a strong s-process component were likely modified by thermal processing during the accretion process. 
Finally, \citet{Dauphas2024} argue for a CI mass fraction for Earth below 10\%, assuming an enstatite chondrite composition for the Earth, but this seems in tension with for example the \ce{^{54}Fe} isotope composition \citep{schiller2020} and the positive \ce{^{100}Ru} isotope anomaly of early Earth \citep{Fischer2020}.}
Taken together, {these findings} illustrate the current debate in understanding the origin and the possible subsequent mixing of the different isotopic reservoirs that are important for understanding mass transport and planet formation in the early Solar System. 

Recently, \cite{liu2022} presented a scenario which proposes that {a temporal evolution of the isotopic composition of the inner disk} is consistent with pebble drift in a viscously expanding protoplanetary disk. Outer-disk pebbles initially move along with the gas outward and only later drift inward as the gas disk gets depleted. 
This causes a significant delay in the arrival of CC material into the inner region of the disk. Particles, limited in size by fragmentation, that are outside $30$\,AU only reach the terrestrial region after 2\,Myr. In this way a too early pollution of the inner NC reservoir is avoided. After this time, {carbonaceous} chondrites form in the outer regions of the disk and the terrestrial planets can accrete their CC signature from inward drifting pebbles. 
A benefit of this scenario is that it allows for Jupiter to migrate inward after starting its accretion in a wide orbit.
{This is} in contrast to the proposed in-situ formation proposed in previous works \citep{kruijer2017, nanne2019}{, where a non-migrating} Jupiter would form an early gap in the gas distribution blocking the drift of solids into the inner disk. \citet{liu2022} argued how this would effectively deplete the NC reservoir quickly, which seems difficult to reconcile with a late-accreted CC imprint to the Earth. 
Alternatively, if the gap is less efficient in blocking inward drifting pebbles, CC material would drift through the gap and mix in the inner disk with the NC solids, in tension with the presence of relatively late-formed ($\sim$$1.5$\,Myr) NC ordinary chondrites. 

It is believed that the different inner and outer isotopic reservoirs must have been present already early in the evolution of the Solar Nebula. One possibility is that the isotopic heterogeneity was passed down from an inefficiently-mixed pre-solar molecular cloud onto the evolving disk \citep{dauphas2002,nanne2019}. 
Specifically, \citet{nanne2019} propose an initial CAI-like disk got polluted by later NC infalling material. {This second component would need to be accreted with low angular momentum to prevent an NC outer disk. In the final step, the authors speculate that an epoch of extreme outward viscous transport may then create a CI-like outer disk that is a mixture of a CAI and NC.} {However, this scenario is difficult to reconcile with numerical simulations showing that gas within an individual prestellar core appears to be very well-mixed \citep{Kuffmeier2016} and that material later accreted onto disk generally has higher angular momentum \citep{huesoguillot2005,kuffmeier2023}. Moreover, gas and solids within the CAI formation zone would be in a disk region rapidly accreting inward \citep{ZhuZ2023}.
}

Alternatively, as we will explore more in this work, a thermal processing event in the disk could have favored the sublimation of the thermally unstable component of the solid grains, creating a relative depletion of  CI-like material in the inner disk and thereby creating the observed dichotomy \citep{trinquier2009,vankooten2016,ek2020,schiller2015}. 

The mechanism most commonly attributed to thermal processing events are the episodic accretion outbursts observed in young T Tauri stars, like FU Orionis.  The luminosity of these young stars can abruptly increase by up to a factor thousand on timescales of tens to hundreds of years \citep{hartmann1996}. During the outburst, disk accretion rates peak at rates as high as $10^{-4}\, \rm M_\odot/yr$ \citep{zhu2007}. 
Observations indicate that there is a tendency for outbursts occurring when disks are massive ($M_{\rm disk} \approx$ $0.011$ -- $0.38$\,M$_\odot$) and have relatively compact radii \citep[$R_{\rm disk} \approx$ $16$ -- $69$\,AU,][]{kospal2021}. 
Nearly all young sun-like stars are estimated to undergo an outburst phase during their Class I and, less frequently, during Class II stages \citep{Scholz2013,contreras2019}.

A wide range of mechanisms have been considered to explain the triggering event in the outbursts. Some internal triggers could include gravitational fragmentation within the disk, followed by the inward migration of the disk fragments \citep{vorobyov2005,Dunham2012}. Alternatively, the activation of the magneto-rotational instability (MRI) in the inner disk, due to gravitational instabilities in the outer disk, could induce the rapid transport of great amounts of material onto the star \citep{zhu2009}. In some cases the outbursts are also thought to be products of external interactions with the late infall of gas from the molecular cloud onto the protoplanetary disk \citep{kuffmeier2018}, or with other stellar companions such as a binary \citep{reipurth2004}. It is not well understood which of these mechanisms dominates the stellar outburst activity, nor how they are potentially related to each other. 

Several works have investigated the implication that stellar outbursts have on the solids present in protoplanetary disks. 
During outbursts the sublimation lines of different volatile species such as water, carbon monoxide and complex organic molecules move significantly outward \citep{Banzatti2015}.
For example, in the case of the 1.3 solar-mass protostar V883 Ori, 
 {an outburst with a stellar luminosity of $L \approx 200$\,L$_\odot$}
displaced the water ice line out to 42--80\,AU \citep{cieza2016,tobin2023}. Displacement of the ice lines also effects the CO ice sublimation and chemistry in the outer disk of young protostars \citep{visser2012,jorgensen2015}. Additionally, more refractory dust can be affected: recent  observations with the James Webb Space Telescope (JWST) of EX Lup strongly argue for post-outburst outward transport of thermally-processed inner-disk dust grains (crystalline forsterite grains) to regions beyond $3$\,AU \citep{Kospal2023}. 
Finally, in the Solar System, stellar outbursts may explain the depletion of refractory carbon in the terrestrial region with respect to cometery values \citep{Binkert2023} and have been linked to the depletion of moderately volatile elements in carbonaceous chondrites \citep{hubbard2014,li2020,Li_2023}.
\citet{johansen_dorn2022} argued that the condensation of solids after an outburst leads to the formation of iron-rich planetesimals in the inner disk, potentially explaining the iron-rich composition of Mercury.

In this study we assume that our Sun underwent outburst events during its formation. The aim is to explore the long-term effect on the solid component after the last major event that thermally processed a significant fraction of the pebble reservoir, while remaining agnostic on the exact mechanism that triggered the outburst in the first place. We focus on the evolution in time of two radially-separated solid species. For simplicity, we will assume the inner solid species is thermally processed and carries the NC imprint. Unfortunately, it is not yet well-understood how exactly thermal processing alters the composition of primordial material by the selective removal of, possibly unknown, isotopically anomalous
carrier species. However, although the precise process is unclear, it does appear from \citet{liu2022} that a large fraction of the material has to be processed, out to distances of 10\,AU soon or during the formation of the protoplanetary disk. Therefore, we consider FU Orionis outbursts that can heat up the disk up to tens of AU and concentrate on the subsequent evolution of the pebble populations in the disk {that were} initially separated by the water ice line. 

The paper is structured as follows.
In \autoref{methods} we describe the initial setup for the disk evolution and pebble population, as well as the considerations taken for the thermal processing of it. We run our simulations for four different initial parameters, described in \autoref{results}. Additionally, in \autoref{implications} we give the implications of our results in the context of the isotopic measurements of chondrites. In \autoref{sec:discussion} we perform additional simulations to test the influence of particle size and briefly discuss the role of Jupiter in possibly reducing inward pebble drift. Finally, we present our conclusions in \autoref{sec:conclusions}.   

\section{Methods}
\label{methods}

\subsection{Gas disk evolution}
\label{sec:methodsgas}
\begin{table*}
\caption{Parameters of the models studied}             
\label{table:1}    
\begin{center}
\begin{tabular}{c c c c c c c c}       
\hline\hline                 
Model & $r_{\rm init}$ [AU]  & $t_{\rm out}$ [yr] & $r_{\rm sep}$ [AU] & $r_{\rm v=0}$ at $t_{\rm out}$ [AU] & $L_{\rm init}$ [$L\odot$]& $L_{\rm out}$ [$L\odot$] & $\alpha$\\    
\hline                        
   \texttt{Nominal}         & 20 & - &  6.90 at $t=1\,\mathrm{kyr}$  & 2.88 at $t=1\,\mathrm{kyr}$ & 1& ~& 10$^{-2}$\\ 
   \texttt{Compact}         & 2.5& - & 7.54 at $t=1\,\mathrm{kyr}$   & 1.62 at $t=1\,\mathrm{kyr}$ & 1&~ & 10$^{-2}$\\
   \texttt{Outburst}        & 20 & $10^4$ & 67.15 at $t_{\rm out}$ & 15.91 at $t_{\rm out}$  &1&1000 & 10$^{-2}$\\
   \texttt{Outburst-NoVisc} & 20 & $10^4$ & 67.15  at $t_{\rm out}$ & 12.53 at $t_{\rm out}$ &1&1000 & 10$^{-2}$\\ 
   \texttt{Med-Outburst} & 20 & $10^4$ & 42.32  at $t_{\rm out}$ & 12.52 at $t_{\rm out}$ &1&500 & 10$^{-2}$\\
   \texttt{Small-Outburst} & 20 & $10^4$ & 14.57  at $t_{\rm out}$ & 12.52 at $t_{\rm out}$ &1&100 & 10$^{-2}$\\
   \texttt{Late-Outburst} & 20 & $10^5$ & 67.15  at $t_{\rm out}$ & 17.42 at $t_{\rm out}$ &1&1000 & 10$^{-2}$\\
   \texttt{Med-visc-alpha} & 20 & $10^4$ & 67.15  at $t_{\rm out}$ & 12.26 at $t_{\rm out}$ &1&1000 & 5$\times$10$^{-2}$\\
   \texttt{Low-visc-alpha} & 20 & $10^4$ & 67.15  at $t_{\rm out}$ & 12.03 at $t_{\rm out}$ &1&1000 & 10$^{-3}$\\
\hline    
\end{tabular}
\end{center}                         
\begin{tablenotes}
    \item \small {The value of $t_{\rm out}$ corresponds to the time when the outburst is introduced in the outburst models. The parameter $r_{\rm sep}$ describes the radial distance where the processed and unprocessed components are separated. The radius $r_{\rm v=0 }$ indicates the initial distance where the gas advection changes direction from moving toward the star to expanding outward viscously to conserve angular momentum.}
\end{tablenotes}
\end{table*}

The time evolution of the surface density of the gas, $\Sigma_{\rm g}$, is given by the diffusion-advection equation
\begin{equation}
    \frac{\partial \Sigma_{\rm g}}{\partial t} = \frac{3}{r}\frac{\partial}{\partial r} \Bigg[r^{1/2}\frac{\partial}{\partial r}\bigg(\nu \Sigma_{\rm g} r^{1/2}\bigg)\Bigg]\,,
\end{equation}
with $r$ the orbital radius \citet{pringle1981}. We adopt the disk viscosity from \citet{shakura1973}, given by
\begin{equation}
    \nu = \alpha c_{\rm s} H\,.
    \label{eq:visc}
\end{equation}
Here, $c_{\rm s}$ is the sound speed, $H$ is the scale height of the gas disk, and $\alpha$ is the viscosity parameter, set to $\alpha=0.01$. 
This high value is motivated by our aim to focus on the earliest stages of disk evolution where disks may be close to being gravitational unstable \citep{Elbakyan2020} and the observed high accretion rates of young disks \citep{Manara2016}. 

\begin{figure}
\centering
\includegraphics[width=8.7cm]{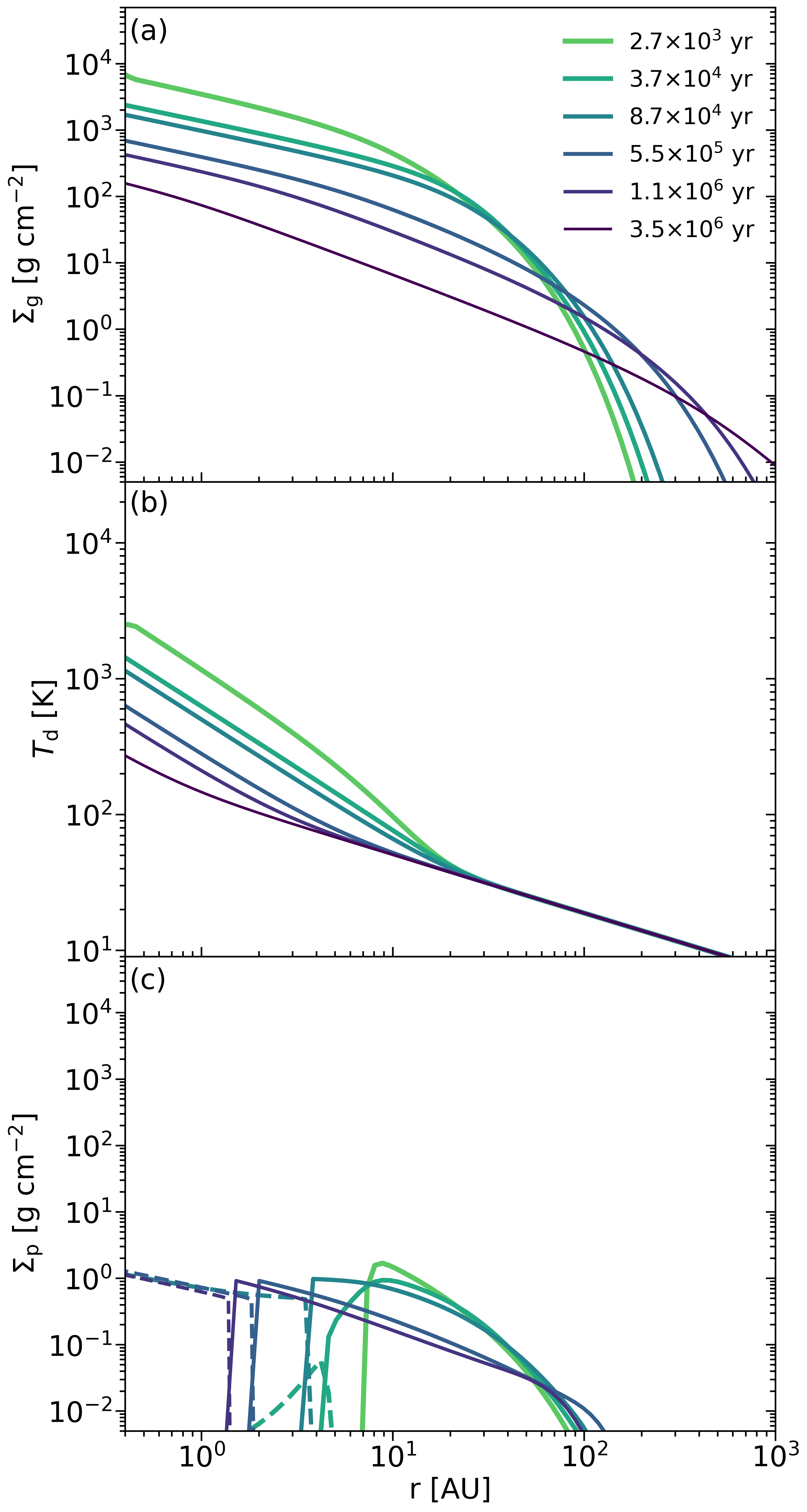}
  \caption{Gas surface density, gas temperature, and pebble surface density evolution for the \texttt{Nominal} disk model. (a) The mass in the inner disk decreases in time due to accretion onto the central star, whereas the mass in the outer gas disk expands into wider orbits to conserve angular momentum. (b) At early times the midplane temperature inside $30\,\rm AU$ is dominated by viscous heating and the outer part of the disk is heated by irradiation from the central star. 
  (c) 
   {
  Solid lines show the distribution of the unprocessed pebbles that are located behind the water ice line (see \autoref{initcomp}). After $\sim$ $1$\,Myr the inner surface density increases since a fraction of these solids moves inward due to radial drift. These pebbles subsequently sublimate at the water evaporation evaporation front now located near approximately $1\,$AU. 
  The surface density of this vapor component is shown with the dashed lines.
  }}
\label{Surface_temp}
\end{figure}

As the initial condition for the gas disk profile, we will take the analytical solution for the surface density of the gas given by
\begin{equation}
    \Sigma_{\rm g} = \frac{M_{\rm d,0}}{2\pi r_{\rm init}^2}\frac{1}{(r/r_{\rm init})}e^{-(r/r_{\rm init})}\,.
\end{equation}
where $M_{\rm d,0}$ represents the initial mass of the disk. 
Here we assumed a power law viscosity $\nu\propto r^\gamma$ with $\gamma=1$, which is valid when using  \autoref{eq:visc} with \autoref{eq:initT} for the initial temperature profile \citep{hartmann1998}. The initial radius, $r_{\rm init}$, is the radius inside of which $\approx$$\,0.6$ of the total mass of the disk is initially located. {To mimic the first stages of disk evolution, we start with a model of a massive and compact disk, which is no longer accreting from the surrounding gas. Our model $t=0$ can thus approximately be placed after the final infall stage and the end of the class I phase of disk evolution.
}
We set the disk mass in all models to $M_{\rm d,0}=0.1M_\odot$ and the range of initial radii considered in our models is listed in \autoref{table:1}. 
These initial conditions for massive and relatively compact disks are broadly consistent with disk observations of FU Orionis objects \citep{kospal2021}. 
We illustrate the evolution of the gas surface density for the \texttt{Nominal} model in panel (a) of ~\autoref{Surface_temp}.

\subsection{Temperature evolution}\label{tempevol}

We calculate the temperature considering the heating produced by viscous turbulence as well as the contribution from the irradiation of the central star.
For the irradiated temperature, $T_{\rm irr}$, we assume that the star is a point-like source as done in \citet{menou2004}, so that the irradiation temperature follows the equation
\begin{equation}
    \sigma_{\rm SB}T_{\rm irr}^4 = \frac{L}{4\pi r^2}\frac{H}{r}\bigg(\frac{\partial \ln H}{\partial \ln r}-1\bigg)\,.
\end{equation}
Here, $\sigma_{\rm SB}$ is the Stefan-Boltzmann constant and $L$ is the luminosity of the central star. 
The term $\big({\partial \ln H}/{\partial \ln r}-1\big)$ represents the shielding factor, where we set  $\partial\ln H/\partial\ln r$ to 9/7 \citep{huesoguillot2005}.
We determine the equilibrium temperature of the viscous heating by balancing the cooling rate
\begin{equation}
    \Lambda_{\rm rad} = \frac{8\sigma_{\rm SB}T_{\rm vis}^4}{3(\tau/2+1/\sqrt{3}+ 1/3\tau)}
\end{equation}
with a heating rate defined by
\begin{equation}
    \Gamma_{\rm vis} = \frac{9}{4}\nu\Sigma\Omega^2\,.
\end{equation}
Here $\tau$ is the optical depth, taken to be $\tau=\kappa \Sigma_{\rm g}$, with $\kappa$ the opacity. Combining both expressions, we find that the temperature component from the viscous turbulence in the disk is given by
\begin{equation}
    T_{\rm vis}^4 = \frac{27\nu \Sigma_{\rm g}\Omega^2}{64 \sigma_{\rm SB}}\bigg(\tau/2+1/\sqrt{3}+1/3\tau\bigg)\,.
\end{equation}
To determine the disk temperature at the midplane, $T_{\rm d}$, we calculate
\begin{equation}
    T_{\rm d}^4 = T_{\rm irr}^4 + T_{\rm vis}^4\,,
\end{equation}
following \cite{kimura2016}. The sound speed relates to the midplane temperature as
\begin{equation}
    c_{\rm s}^2 = \frac{k_{\rm B} T_{\rm d}}{\mu m_{\rm H}}\,.
\end{equation}
Here, $\mu=2.34$ is the mean molecular weight, and $m_{\rm H}$ the mass of the hydrogen atom. 

We assume a simple opacity prescription, as we keep the opacity fixed at $\kappa=1\,\rm cm^2/g$. We implement a maximum allowed value in the temperature distribution, which we set to $T_{\rm max}= 2500$\,K. This is because temperatures in the inner disk are unlikely to exceed a few $10^3\, \rm K$ as solids sublimates and gas opacities start dominating \citep{birnstiel2010,li2021a}.

When initiating the simulations, we adopt an initial temperature profile as described in \citet{hayashi1981},
\begin{equation}
    T_{\rm d} =  280\,\bigg(\frac{r}{\mathrm{AU}}\bigg)^{-1/2}\,\mathrm{K}\,.
    \label{eq:initT}
\end{equation}

An example of the evolution of the temperature profile including contributions from both irradiation and viscosity
is shown in panel (b) of \autoref{Surface_temp}.

For those models where we include an accretion outburst (see \autoref{sec:outburst}), we modify the luminosity after the disk has converged to thermal equilibrium. 
Mass accretion episodes are accompanied by a rise in the accretion luminosity in the innermost disk, which effectively increases the heating from irradiation exerted onto the outer disk \citep{zhu2007}. 
We do not explicitly model the accretion outburst, but instead alter the stellar luminosity by adding the additional accretion luminosity, in line with other theoretical studies \citep{Binkert2023,Houge2023}.
In practice, we alter the luminosity from  $L_{\rm init} = 1\,L_\odot$ to $L_{\rm out} = 1000\,L_\odot$ in the nominal outburst case (see \autoref{table:1}). 
The outburst is introduced after a time $t_\mathrm{out} = 10^4\,\mathrm{yr}$, in line with an outburst recurrence of $\approx$100\,kyr in Class II young stellar objects \citep{contreras2019}, and lasts for $100$\,yr \citep[and references therein]{fischer2023}. 
This setup results in an effective water iceline excursion out to $\sim$$70$\,AU, in line with the observed thermal expansion seen in V883 Ori \citep{tobin2023}. {We will also explore different outburst timings \autoref{sec:outburst} and lower luminosities around of $100$--$400$\, $L_\odot$ \citep{audard2014} in \autoref{app:outbursttime}. For simplicity, we will only consider the last major outburst, meaning that any imprint from previous smaller outbursts is lost \citep{Li_2023}.
}

\subsection{Solid component evolution}
We will consider several pebble populations in the disk that are individually evolved according to the advection equation
\begin{equation}
    \frac{\partial \Sigma_{\rm p}}{\partial t}=-\frac{1}{r}\frac{\partial}{\partial r} (r\Sigma_{\rm p}\varv_r)\,.
\end{equation}
Here $\Sigma_{\rm p}$ is the surface density of the solid component made out of pebbles, $r$ the orbital radius and $\varv_r$ the radial velocity of the solid particles. This velocity has contributions from the radial movement of gas, $\varv_{\rm gas}$, and from drag-induced radial drift, $\varv_{\rm drift}$. The radial gas velocity is given by
\begin{equation}
    \varv_{\rm r, gas} = - \frac{3}{\Sigma_{\rm g} r^{1/2}}\frac{\partial}{\partial r}\bigg(\Sigma_{\rm g}\nu r^{1/2}\bigg)\,,
\end{equation}
and the drag-induced drifting velocity for the pebbles is given by \citep{weidensch1977} 
\begin{equation}
    \varv_{\rm r, drift} = -\frac{2\varv_n}{\text{St}+\text{St}^{-1}}\,.
    \label{eq:drift}
\end{equation}
Here, the Stokes number $\text{St}$ is given by
\begin{equation}
    \text{St} = \frac{\pi}{2}\frac{\rho_{\rm s} R_{\rm p}}{\Sigma_{\rm g}}
\end{equation}
in the Epstein drag regime, with $R_{\rm p}$ being the particle size and $\rho_{\rm s}$ its density.  {
Initially, we consider a constant particle size of $R_{\rm p}=0.1$\,mm, consistent with the lower bounds on particle sizes inferred from the outer parts of observed protoplanetary disks \citep[e.g.,][]{Perez2015,Macias2021,Ohashi2023}. 
We explore the influence of particle size in \autoref{sec:parsize}.
} Finally, the sub-Keplerian speed of the gas $\varv_{n}$ \citep{brauer2008}, is given by
\begin{equation}
    \varv_{\rm n} = - \frac{1}{2}\frac{\partial \ln P}{\partial \ln r} \bigg(\frac{H}{r}\bigg)^2 \varv_{\rm K}\,,
\end{equation}
with $\varv_{\rm K}$ the Keplerian velocity and $P$ the pressure of the gas.
 {
Panel (c) of ~\autoref{Surface_temp} shows the evolution of the solid surface density in the the \texttt{Nominal} model, for the population of unprocessed solids (see \autoref{initcomp}).

}

We do not include here additional turbulence-driven radial particle diffusion. The degree of this contribution is dependent on the physical source of turbulence, which is different than the alpha-viscosity used here only for the secular evolution of the gas disk. For example, turbulence may be highly anisotropic and largely suppressed in the dense pebble midplane layer, as is the case for {turbulent motions triggered by the vertical shear instability} \citep{Flock2017,schafer2022}.
{
Furthermore, observations hint that particle stirring is low in protoplanetary disks \citep{Jiang2023,Zagaria2023}. 
}
Thus, our work can be seen as exploring the case of minimal particle stirring, while future works could explore physical-driven additional radial particle diffusion.

\subsection{Initial populations and sublimation} \label{initcomp}

{We will explore how the early thermal processing influences the composition of pebbles and how their movement affects the final compositions of planetary and meteoritic bodies in the disk. For simplicity, we assume two different pebble populations: unprocessed and processed pebbles (indicated with subscripts $up$ and $p$ henceforth). }

\begin{itemize}
    \item Unprocessed (up) material has never experienced temperatures above the water sublimation point. We take a fixed sublimation temperature for water $T_{ \rm sub,up}=160\, \rm K$ \citep{ros2013}.
    \item Processed (p) solids have experienced temperatures between the water sublimation temperature and the temperature at which solid refractory materials get altered. For this latter temperature, we take the sublimation temperature of  {a representative refractory element}, Si, with $T_{50, \rm p} = 1310\, \rm K$. 
    Here $T_{\rm 50}$ stands for the 50\% condensation temperature, where half of a given element is found in the solid state and the other half in the gas, for a solar gas mixture \citep{wood2019}.
    
\end{itemize}

The {two} populations considered can be found in one of two states in the disk: gas or solid form. In the gas form the population follows the same dynamic behavior as the gas disk, whilst the solid component experiences radial pebble drift. When unprocessed solid material crosses the sublimation line, it gets fully sublimated at a rate
\begin{equation}
    \bigg(\frac{\partial \Sigma_{\rm d, up}}{\partial t}\bigg)_{ \rm sub} = f_{\rm sub,up} \epsilon \Omega \Sigma_{\rm d,up} .
    \label{eq:sub_rate}
\end{equation}

\noindent The term $ \Sigma_{\rm d, up}$ is the surface density of the unprocessed solid component. Here $\epsilon$ is an efficiency parameter that regulates the rate of material sublimated.
We simply set $\epsilon = 1$, corresponding to sublimation on orbital timescales. When considering the desorption of monolayers of water, \citet{piso2015} argued for lower efficiencies exceeding $\epsilon = 10^{-2}$ for sub-mm-sized grains. Our results are not sensitive to the precise choice, but long sublimation timescales, and the possible disintegration of grains, could matter in the case of short-timescale luminosity outbursts (for more see \autoref{sec:parsize}).
Finally, the term $f_{\rm sub,up}$ describes the mass fraction that gets sublimated. We consider that for $up$ pebbles crossing the water sublimation line, half of the pebble mass fraction ($f_{\rm sub,up} =0.5$) is added to the $p$ solid component. This is an ad hoc value for the amount of volatile loss, corresponding approximately to the water mass fraction of a solar composition. Smaller water mass fractions of around a third have been inferred for cometary solids \citep{Fulle2019}, but on the other hand, less volatile species could also be lost as pebbles disintegrate.

\begin{figure*}[!t]
   \centering
   \includegraphics[width=16cm]{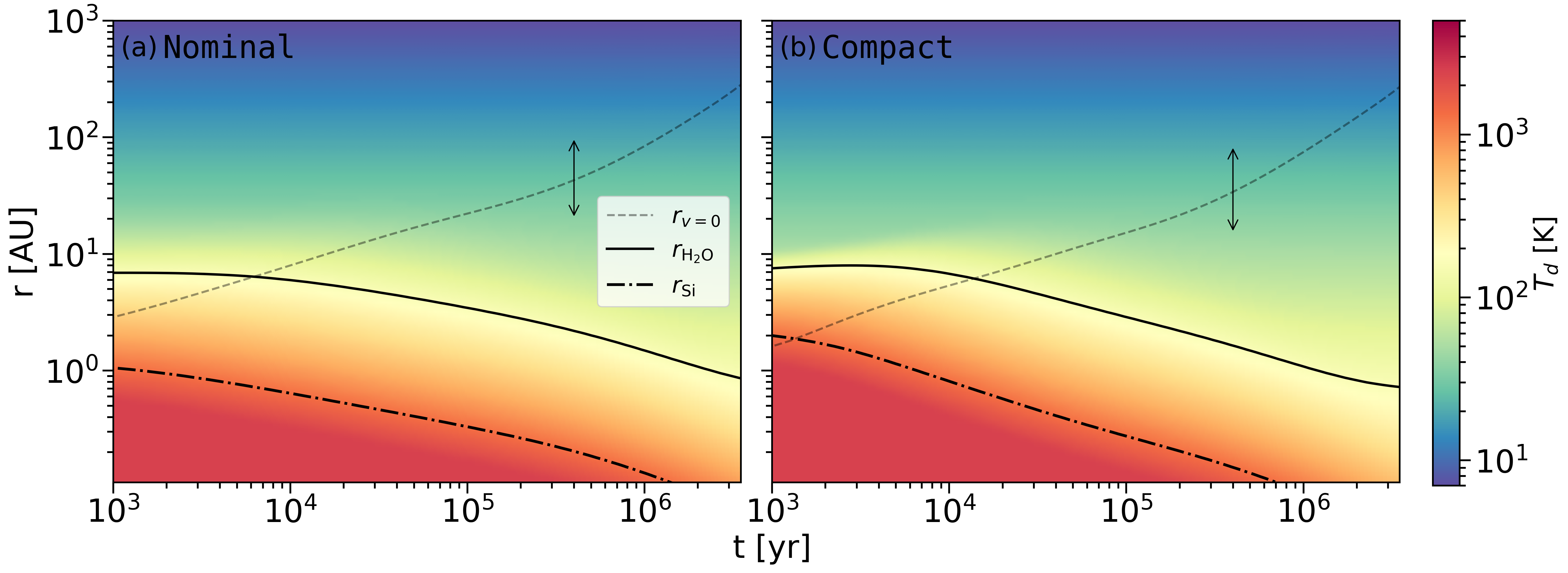}
   \caption{Temperature structure evolution for the (a) \texttt{Nominal} and (b) \texttt{Compact} cases. The color-bar shows the disk temperature at the midplane. The {dashed gray} line shows the location where the gas advection changes direction: the material inside the line is accreted onto the star and the material outside is moved outward due to the viscous expansion of the gas disk. The {solid and dash-dotted lines correspond to the location of the water and silicates sublimation lines with temperatures of $160\rm \,K$ and $1314\rm \,K$, respectively.}}
    \label{Heatmap1}
\end{figure*}

The processed material in the disk starts with all the material that has experienced $T> T_{\rm sub,H2O}$ during disk formation and is subsequently complemented by material that gets heated above that temperature during disk evolution. 
We do not consider recondensation of $up$ vapor to $up$ solid, if $up$ vapor is advected outward across its sublimation line into a cooler zone in the disk.
We found that this choice has little impact on our models and it allows for an easier interpretation of the results. With this model choice all $up$ material are pristine solids that have never been processed during the full disk lifetime. Moreover, the actual process is complicated as condensation prefers nucleation onto small grains and is grain composition dependent \citep{ros2013}. We, on the other hand, trace the solid component only via the larger pebbles that make up most of the solid mass. {The processed $p$ solids fully sublimate when crossing the silicate sublimation line. It remains only as a vapor component  {that is unable to re-condense onto the solid component, as explained above}.} 

In practice, we set up the initial condition for the different pebble species in the following way. We start evolving the pebbles after $1\,$kyr when the irradiation and viscous heating have reached equilibrium in the disk and the sublimation lines are at maximal radial separation. To set the initial separation radius between the processed and unprocessed solid components we find the location in the disk where the temperature matches the $T_{\rm 50}$ for water and define it as the separation radius, where everything outside it is considered to be unprocessed pebbles and inside it is processed pebbles. 

\section{Results}
\label{results}

{We run simulations for a total of {nine} models with parameters that are given in \autoref{table:1}.} First,  {in \autoref{compactresults}}, we consider two models with disks that have heating contributions from both irradiation and viscosity, but differ in their initial radius $r_{\rm init}$. The first model, the \texttt{Nominal} model, has an initial radius of $r_{\rm init} = \rm 20\,AU$, while the more \texttt{Compact} model has a smaller initial radius of $r_{\rm init} = \rm 2.5\,AU$. 
This latter model is inspired by models of ultra-compact disk formation \citep{Hennebelle2020,morbidelli2022}, which may be difficult to reconcile with views of a more long-term disk build up, in part driven by streamers \citep{valdivia2022,kuffmeier2023}.

{In the following \autoref{sec:outburst}, we consider two models where we follow the evolution after an accretion outburst. In model \texttt{Outburst}, we include viscous heating, but we also show a case without viscous heating (\texttt{Outburst-NoVisc}) and varying outburst intensities (\texttt{Small-outburst} and \texttt{Med-outburst}). Finally, we also include three additional models to explore the other parameters. We modify the time at which we introduce the outburst in \autoref{app:outbursttime} (model \texttt{Late-Outburst}), and the amount of viscosity in \autoref{app:alphavar} (\texttt{Low-visc-alpha} and \texttt{Med-visc-alpha}).}

\subsection{Nominal and compact models} 
\label{compactresults}

\subsubsection{Evolution of the disk temperature}

The decrease of the surface density decreases the viscous heating, which cools the disk and makes the sublimation lines move inward. \autoref{Heatmap1}, panel (a), shows how sublimation lines move inward in the \texttt{Nominal} model as the surface density decreases in the region with inward accretion. This latter region is marked with the $r_{v=0}$-line, which describes the position where material changes advection direction from moving inward inside of the line to moving outward outside of it. 

In the \texttt{Compact} model a higher mass fraction of the initial disk mass is located in the inner disk with more substantial viscous heating, allowing more than 70\% of the solid mass fraction of the whole disk to be thermally processed, compared to 15\% in the \texttt{Nominal} case. {The silicates line is more distant from the star, at $\sim$$2\,\rm AU$, as shown in panel (b) of  \autoref{Heatmap1},} due to increased viscous heating in the very inner disk. However the outer disk is less affected and therefore the initial location of the water sublimation line is located similarly around $7\,\rm AU$ for this more compact disk. The $r_{\rm v=0}$ line has a similar behavior in both cases, starting within $2\,\rm AU$ and moving outward as the disk expands with time.

\begin{figure*}[t!]
   \centering
   \includegraphics[width=17cm]{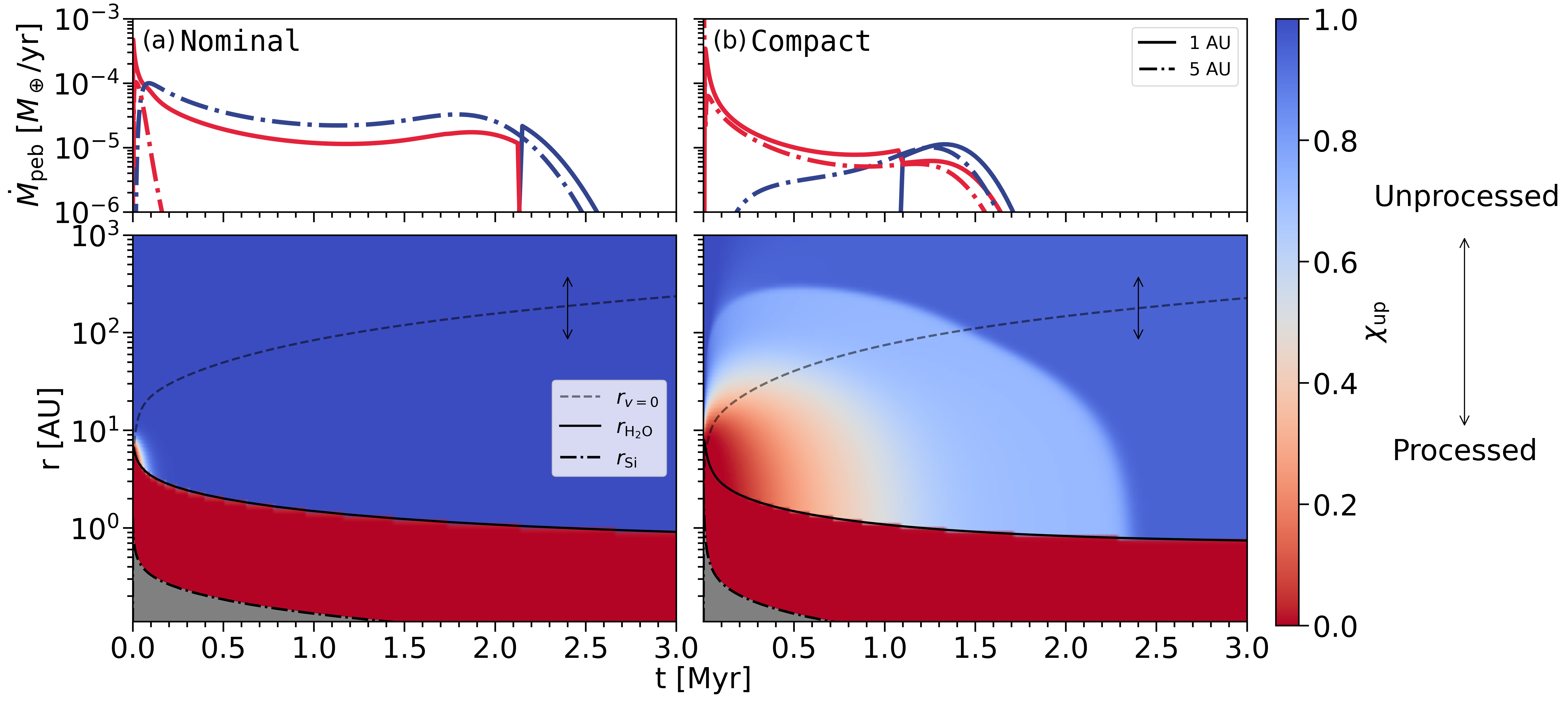}
   \caption{
   Time evolution of the distribution of processed and unprocessed pebbles, for the \texttt{Nominal} case (left panels a) and \texttt{Compact} case (right panels b).
   The top plots show the pebble flux evolution for the processed (red) and unprocessed (blue) pebble components at two different radii (full line $1$\,AU and dashed line $5$\,AU).
   The bottom panels illustrate the evolution of the mass fraction of unprocessed pebbles, following the color code of the legend, at different radii and times in the disk.
   In the \texttt{Compact} case, the processed component is more efficiently  advected outward initially compared to the \texttt{Nominal} case. This results in more mixing with the unprocessed pebbles, but also in loosing most pebbles from the disk within already $2$\,Myr.}
    \label{Conmap1}
\end{figure*}

\subsubsection{Pebble fluxes and mass fraction of unprocessed material}
\label{sec:xupsection}

We now focus on studying the effects of advection and drift in the evolution of the {two} pebble populations. We show the evolution of the absolute value of the pebble flux in \autoref{Conmap1}. The top panels show the pebble fluxes at different radii for both the $up$ and $p$ components as a function of time. The bottom plots show the evolution of the mass fraction of unprocessed pebbles in relation to the total mass of solids present throughout the disk, 
\begin{equation}
    \chi_{\rm up} = \frac{\Sigma_{\rm up}}{\Sigma_{\rm up}+\Sigma_{\rm p}}\,.
\end{equation}
In the \texttt{Nominal} model, panel (a), the processed component is constantly replenished by inward drifting unprocessed material across the water sublimation line. In this case, the $r_{v=0}$ line lays far outside of the water sublimation line, so that little processed material mixes with outer disk material. The top panel of \autoref{Conmap1} shows that the unprocessed component becomes dominant at $1\,$AU after {$2.1\,$Myr} and final pebbles are almost pure $up$.  

For the \texttt{Compact} model in panel (b), more processed material moves outward due to the initial location of the $r_{v=0}$ well within the water ice line, causing the outer disk to get more enriched with $p$ pebbles, compared to the \texttt{Nominal} model. However, pebbles are lost quickly compared to the \texttt{Nominal} case, due to the small radial extent of the disk. After $\sim$$2$ Myr  there is no material left to accrete.

We conclude that neither of these scenarios are able to explain the presence of processed material outside the water ice line at late times. In the \texttt{Nominal} case, the terrestrial region outside $1\,\rm AU$ is dominated by the icy pristine $up$ pebbles from the outer disk, because of the inefficient outward mixing of early processed material. We will argue in \autoref{implications} that this is in apparent conflict with the isotopic composition of meteorites, in the sense that they do not show an exclusive $up$-like composition, which would then correspond to a CI-like isotopic imprint. 

In the compact case, more material {is} processed at early times and diffused outward. However, even using a high viscosity ($\alpha=10^{-2}$), which promotes outward expansion, the pebble flux dries up well before the inferred formation times of the terrestrial planets and the undifferentiated minor bodies of the asteroid belt (this is further explored in \autoref{implications}). We are thus faced with the need for an alternative model where the separation of the populations is located well in the outward moving portion of the disk so that the processed material can be efficiently pushed outward.

\subsection{Outburst models}
\label{sec:outburst}

FU Orionis outbursts are known to be present in the early stages of disk evolution and play a fundamental role in increasing the temperature in the outer disk, therefore we study their influence in the thermal processing of solids in the disk. Although young disks are understood to undergo repeated outbursts, we here limit ourselves to modeling the last major outburst. We first introduce an outburst into a disk similar to the \texttt{Nominal} model described in \autoref{tempevol}. {Then, we remove the viscous heating component,  as the magnitude of this contribution is difficult to precisely determine due to its opacity dependence and assumption on vigorous MRI-driven turbulence \citep{Nomura2002,Bitsch2015}. Under nonideal magnetohydrodynamic conditions, viscous midplane heating may be  largely suppressed \citep{Mori2019,mori2021}.} {Additionally, because the range of observed luminosities in outbursts is very wide, displaying changes between $10$ and $1000$\,L$_\odot$ \citep{hartmann1996,audard2014,fischer2023}, we include an exploration of this parameter, comparing models \texttt{Small-outburst}, \texttt{Med-outburst} and \texttt{Outburst-NoVisc} with three different luminosities that cover the higher-end of the observed systems. We also explore the timing of the outburst in \autoref{app:outbursttime}.}

\subsubsection{Evolution of the disk temperature}

\begin{figure*}[!ht]
   \centering
   \includegraphics[width=16cm]{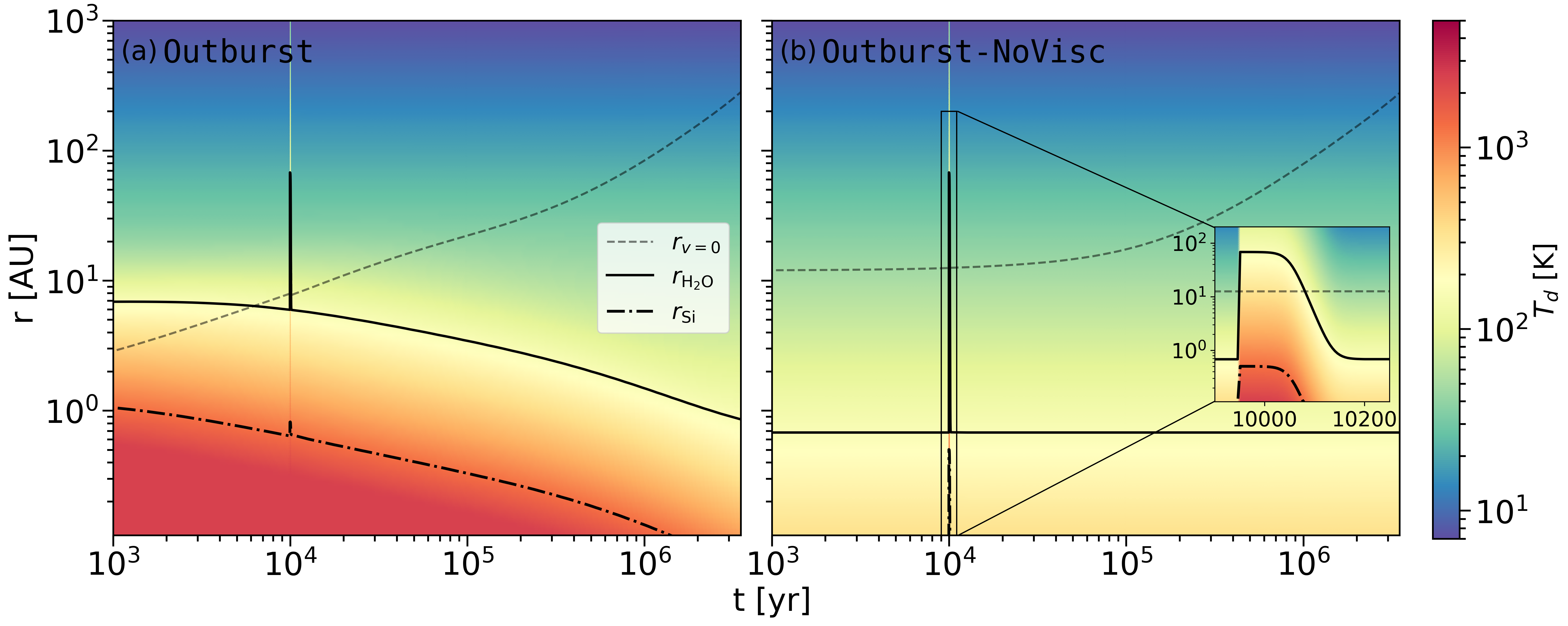}
   \caption{Temperature evolution for disks that include a luminosity outburst of $1000\,L_\odot$. Both cases have an initial disk radius of 20 AU. Panel (a) shows a disk that includes viscous heating and case (b) shows a disk that only has heating from irradiation. In the disk of the $Outburst$ model (panel a) the sublimation lines move slowly inward due to the contribution from viscous heating which eventually decreases, but the material in the inner disk of the \texttt{Outburst-NoVisc} model (panel b) is too cold for the {silicate lines} to be present.}
    \label{Heatmap2}
\end{figure*}

{We compare the temperature evolution of two outburst cases (\texttt{Outburst} and \texttt{Outburst-NoVisc}) in  \autoref{Heatmap2}}. Both panels show the same general behavior in sublimation line evolution; the sudden increase in temperature at the beginning of the outburst affects mostly the outer region of the disk. The water sublimation line moves its position drastically out to $\sim$$70\,\rm AU$, while the inner sublimation lines only undergo a minor excursion. However, there are also significant differences between the two models. In model \texttt{Outburst}, the $r_{v=0}$ line is initially located within $0.5\, \rm AU$.  {Subsequently, the $r_{v=0}$ line  quickly moves to $10\,\rm AU$ in the first $10$\,kyr of evolution as seen in panel (a) of \autoref{Heatmap2},while in the \texttt{Outburst-NoVisc} run the  $r_{v=0}$ line remains stable at approximately 10\,AU during this period.}
In the latter case, the separation between inward and outward gas advection is located at $r_{v=0} \approx r_{\rm init}/2$, as expected in a disk heated by irradiation only \citep{hartmann1998,liu2022}. However, as seen here, in disks with additional viscous heating this relation, which assumes a power-law viscosity profile, does not hold longer. The increased viscosity in the inner disk drives $r_{v=0}$ closer to the host star. 
Another difference is that without any viscous heating the water ice line is located well within the terrestrial region at $0.8\,\rm AU$.
We note that the \texttt{Outburst-NoVisc} model is the only case in our study where, due the lack of viscous heating, the inner disk ($<$$0.5\,\rm AU$) is heated above $1000\,\rm K$ only for a small fraction of time ($\sim$$100\, \rm yr$). Lastly, the temperatures outside $\sim$$10\,$AU before and after the outburst remain similar in time throughout all four cases, as they are regulated by nominal stellar irradiation.

\begin{figure}[!ht]
   \centering
   \includegraphics[width=9cm]{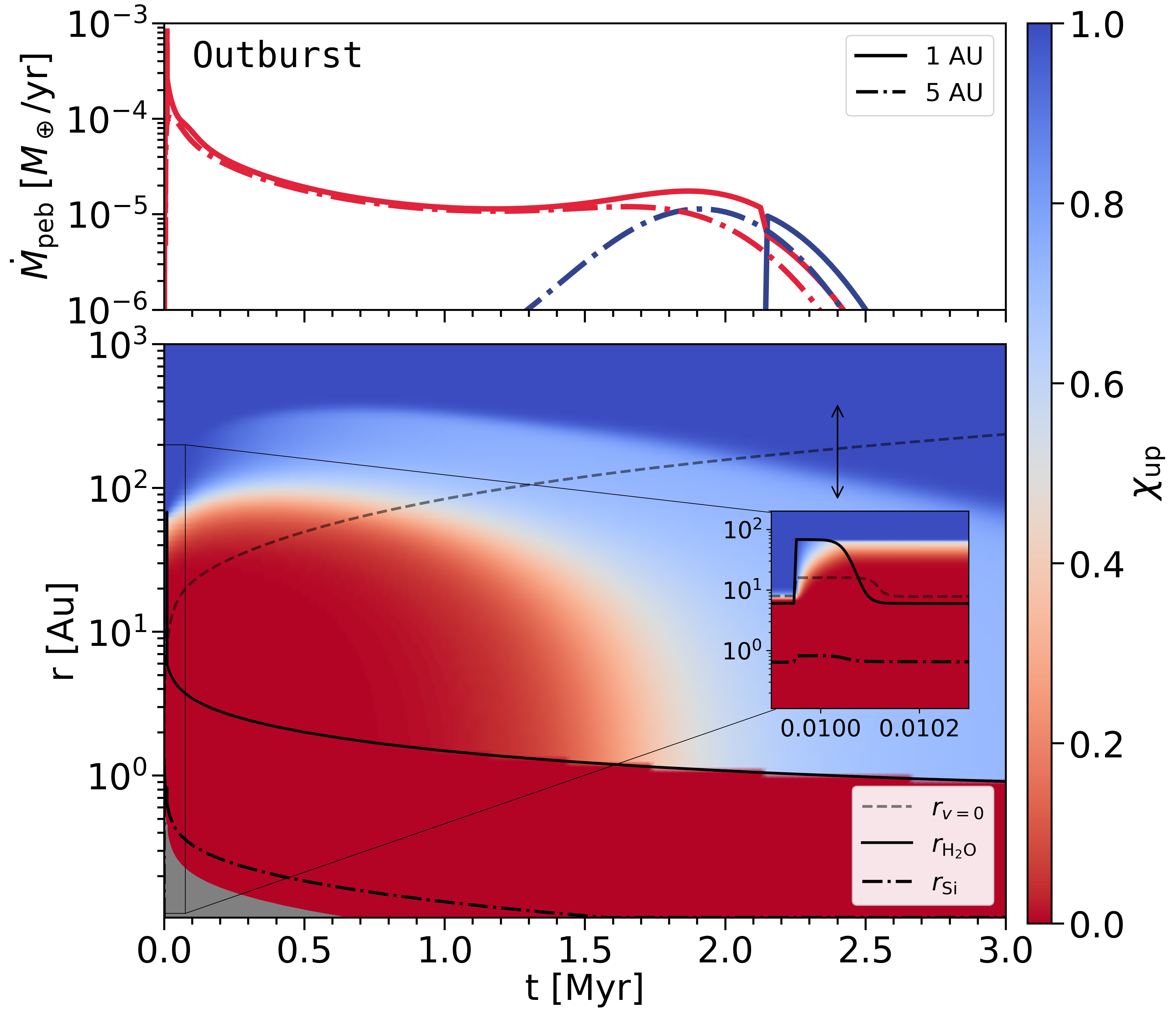}
   \caption{{Same as \autoref{Conmap1}, but for the \texttt{Outbust} model. The location of the water sublimation line, which separates the processed and unprocessed pebbles and is indicated by the solid line, is located beyond the location where the gas velocity reverses ($r_{\rm H_2O}>r_{v=0}$), indicated by the dashed line, at the time of the outburst. As a results we see substantial mixing of both up- and p-pebbles even at late times.}}
    \label{outburst_map}
\end{figure}

\subsubsection{Pebble fluxes and mass fraction of unprocessed material}

{We now analyze the effect of the outburst on the composition of the pebble fluxes. For the \texttt{Outburst} model ($L=1000\,L\odot$), we can see in the upper panel of \autoref{outburst_map} that the processed pebble component dominates the pebble flux until the water sublimation line crosses $1\,$AU at $2.1\,\rm Myr$.} The bottom panel also shows a high degree of mixing outside the water sublimation line, making the composition $p$-pebble dominated out to $100\,\rm AU$ within the first Myr of evolution. This is because the separation radius between the two populations, the water ice line, has been pushed outward due to the increase in heating caused by the outburst. This therefore results in a large {mass} fraction of 40\% of all processed material being located in the outward-moving section of the disk, causing $p$-pebbles to advect outward into the unprocessed reservoir. 

\begin{figure*}[!ht]
   \centering
   \includegraphics[width=18cm]{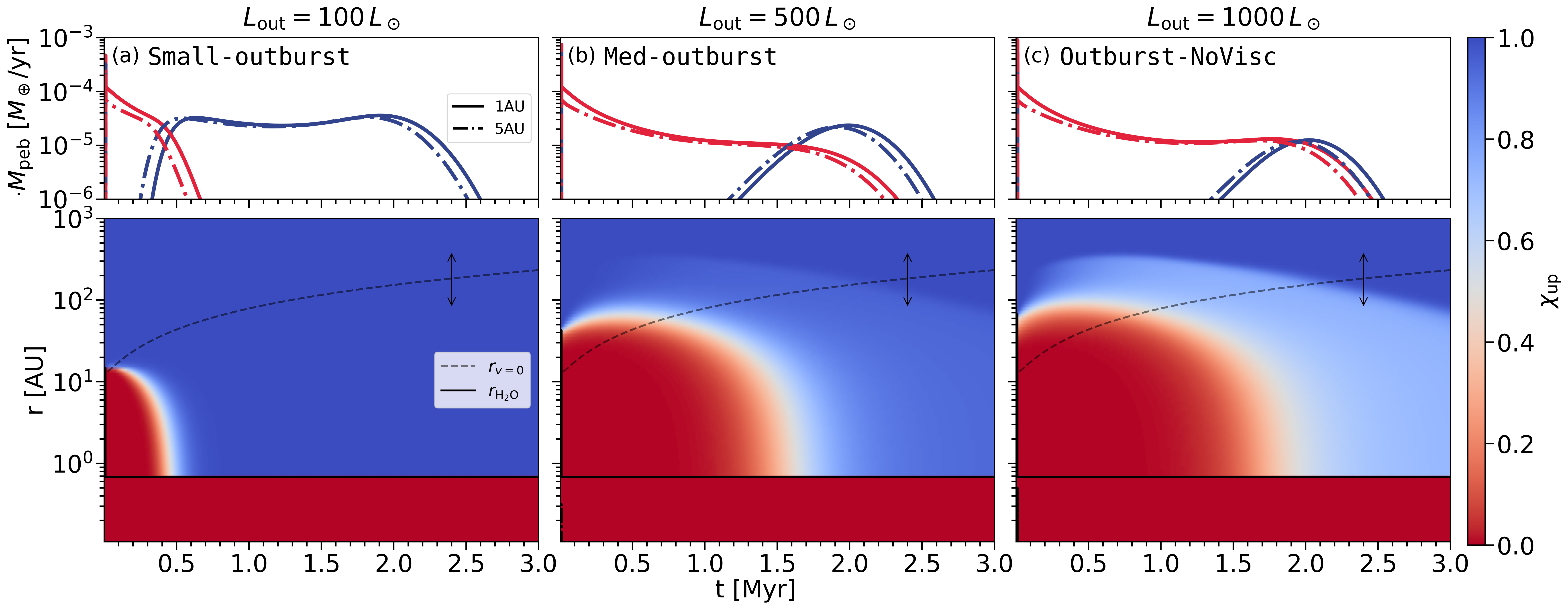}
   \caption{{Same as \autoref{Conmap1}, but for the outburst models that do not include viscous heating (\texttt{Small-outburst}, \texttt{Med-outburst} and \texttt{Outburst-NoVisc}, respectively). }}
    \label{fig:all_outburst}
\end{figure*}

{In \autoref{fig:all_outburst}, we show the evolution of the pebble fluxes for models without viscous heating, but with outbursts that reach up to $L=100\,L_\odot$, $500\,L_\odot$ and $1000\,L_\odot$. They show a similar behavior in terms of unprocessed-processed mixing, with an initial over-abundance of $p$-component that gets taken over by unprocessed pebbles. The main difference between these models is how far away the iceline gets pushed outward, which in return affects how much material gets processed and the time at which the transition from $p$ to $up$ occurs. The heating of the disk during a strong ($L=1000\,L_\odot$) outburst pushes the separation radius of the populations outward and proves to be an efficient way of inducing mixing in the early stages of the system. This, in turn, allows the late delivery of unprocessed material into the inner disk. This stands in contrast to the disks with weaker outbursts, where not enough material is moved out, creating an early delivery of unprocessed materials after 0.5 Myr and 1.5 Myr in the \texttt{Small-outburst} and  \texttt{Med-outburst} models, respectively, as seen in panels (a) and (b) of \autoref{fig:all_outburst}. }

\section{Application: an evolving nucleosynthetic composition}\label{implications}

\begin{figure}[!h]
   \centering
   \includegraphics[width=9cm]{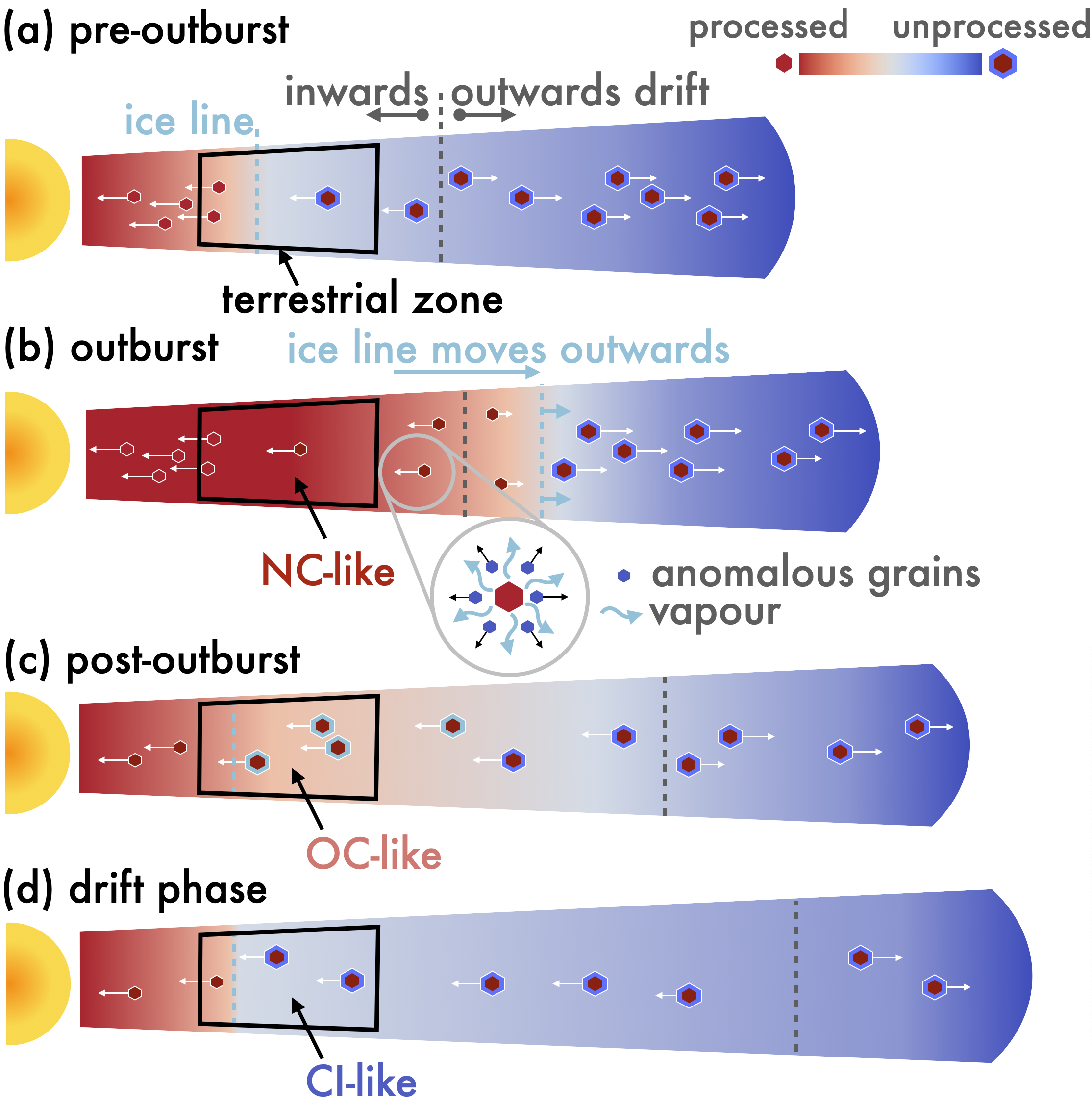}
   \caption{
   {Thermal processing of icy grains. Panel (a) shows the state of the disk before a stellar outburst. The blue dash line represents the water ice line, the gray dashed line the radial location ($r_{v=0}$) that separates the inward- and outward-drifting pebbles, and the black-outlined region the terrestrial planet-forming region.  During an outburst (panel b) the ice line moves outward and  particles in the inner disk loose their icy mantles containing isotopically anomalous grains. Afterwards, the disk cools down and the iceline retreats (panel c) resulting in the inner disk with a mixture of processed and unprocessed grains, altering the solid composition from NC to OC like. However, on longer timescales evermore unprocessed pebbles drift inward resulting in a pebble composition that is close to CI outside the water ice line (panel d).}
   }
\label{cartoon4steps}
\end{figure}

Meteorites sample the composition of the pebble flux at different locations and times during the evolution of the Solar Nebula. 
This is because their parent bodies are believed to form after dense pebble swarms collapse on orbital timescales, triggered by the streaming instability \citep{Youdin2005,johansen2007}. 
Here, we assess if our model for the pebble flux composition can match the evolution of the nucleosynthetic composition of different meteorite classes with different accretion ages.
In order to do so, we will consider that the carriers of the isotopic differences between different meteorites are small presolar grains. We then make the key assumption that these grains are (partly) irreversibly lost during the sublimation of the icy shells on dust particles that enter inwards of the water iceline during disk evolution (\autoref{cartoon4steps}). We speculate here that, once small nanoparticles are lost from the host pebble, they are trapped below the electrostatic barrier \citep{Okuzumi2009} and do not efficiently re-coagulate \citep{Matthews2012,Akimkin2023}. 
We further discuss this assumption in \autoref{sec:parsize}.

Currently, it is not well-understood which types of presolar grains are the carriers of different isotopic anomalies and how they are distributed within the precursor solids \citep[for reviews, see][]{Nittler2016,Hoppe2022}. 
These very small grains, nanometers to micrometers in size, can still be identified in fine-grained matrix \citep{floss2016} and ultra-primitive clasts \citep{Nguyen2023}.
However, these grains are easily destroyed in more thermally-altered materials by aqueous alteration, thermal metamorphism or prolonged oxidation \citep{Davidson2014}. 
It is therefore complex to establish stardust abundance differences in matrix, chondrule and CAI {precursor} materials and it is difficult to explain the differing isotopic compositions by solely different matrix mass fractions \citep[][see also \autoref{sec:companalysis}]{Qin2010}.
Nevertheless, recent Ryugu data appears consistent with the notion that inner disk material is depleted in isotopic-anomalous stardust compared to primitive clasts that are likely tracing primitive outer disk solids \citep{Nguyen2023}.

{
In what follows we will not explicitly model the formation and possible outward viscous transport of the isotopically anomalous CAI/AOA that likely formed early in the evolution of the disk close to the host star \citep{Connelly2012,Larsen2020}. They are unlikely to be the sole carriers of the observed isotopic spread, due to their non-chondritic composition \citep[e.g.,][]{nanne2019,vankooten2021,Hellmann2023}, but their role in possible causing the observed spread of the CC group in isotope space will be discussed in \autoref{sec:companalysis}.}
{We will neither explicitly model the formation and evolution of the highly-processed chondrules found in the matrix that makes up chondrites that also show an NC-CC spread \citep{Schneider2020}. }

\begin{table*}[!ht]
    \tabcolsep=0.11cm 
    \centering 
\input{table.tex}
\end{table*}

\subsection{Different meteoritic classes and selected isotopes}
\label{subsec:introclass} 
Meteorite ages can be conceivably constrained thanks to the presence of short-lived radionuclides \citep{Davis2022} or long-lived chronometric systems \citep{Connelly2017}.
Dating constraints used here are obtained with different methods;  
the \ce{^{82}Hf}--\ce{^{182}W} isotope system for iron meteorite parent bodies \citep{Kruijer2013, Kleine2009}, 
the lower-limit aqueous alteration ages of chondrites from radiometric \ce{^{53}Mn}--\ce{^{53}Cr} \citep{doyle2015}, and upper-limit chondrule ages from \ce{^{26}Al}--\ce{^{26}Mg} \citep{Pape2019} or U-corrected Pb-Pb dating \citep{Bollard2017}, see \autoref{tab:nucleosyn} for more.
In contrast, the primordial formation locations of meteorite parent bodies prior to their possible dynamical delivery to the asteroid belt, are much more enigmatic (see introduction). 
Below, we briefly review key characteristics of the different meteorite classes we consider here.

\begin{itemize} 
    \item 
    {The carbonaceous chondrites consist of undifferentiated chondrites (CI, CO, CV, CM, CR) that} have seen little thermal processing \citep{scott2014} and have an elevated water content compared to other chondrites \citep{alexander2012}. 
    The CI-class is special in being mainly composed out of fine-grained matrix material that has a solar-like chemical composition, whereas other carbonaceous chondrites also contain variable abundances of chondrules, metal and calcium-aluminum inclusions \citep{scott2014}. \autoref{tab:nucleosyn} lists the accretion ages of the different 
    {carbonaceous chondrites considered here}. 
    In general, the carbonaceous chondrites formed relatively late, approximately 2-5\,Myr after CAI formation. Interestingly, the samples returned from asteroid Ryugu are CI-like, as probed via their iron isotopic composition \citep{Hopp2022Science}.
    \item 
    The non-carbonaceous chondrites are composed of the enstatite (EC), Rumuruti chondrites (RC), and ordinary chondrites (OC) that have a distinct chemical and isotopic composition from the {carbonaceous chondrites}.
    They are reported to have formed earlier, within the first 2\,Myr of disk evolution (\autoref{tab:nucleosyn}) and to have seen a higher degree of thermal processing \citep{trinquier2009, larsen2011, Miyazaki2021}.
    
    \item Achondrites originate from differentiated parent bodies that formed early in the evolution of the protoplanetary disk ($\lesssim 1.5$\,Myr, see \autoref{tab:nucleosyn}). 
    The parent bodies of these meteorites are unknown, with the exception of the HED meteorites that originate from Vesta \citep{McSween2011}. The early-formed ureilite class are believed to sample some of the the earliest inner-disk solid materials \citep{schiller2018}.
     {
    The iron meteorite group represents differentiated Fe-Ni cores of distinct planetesimals \citep{Burbine2002}. The complete melting of the iron meteorite parent bodies may imply even earlier formation, particularly if the 
    \ce{^{26}Al}/\ce{^{27}Al} was lower than the nominal value \citep{larsen2011,johansen2023}.
    }
\end{itemize}
{
In summary, a timeline is established where achondrites ($\lesssim 1.5$\,Myr) formed before the non-carbonaceous chondrites ($\lesssim 2$\,Myr), and where the carbonaceous chondrites generally have the oldest ages ($\gtrsim 2$\,Myr). 
However, determining a temporal sequence \emph{within} these groups is more difficult \citep{Spitzer2020}: accretion age estimates come with significant spread, driven by formal measurement error, different methods, and modeling assumptions in the case of achondrite accretion ages \citep{kruijer2017,Spitzer2021}.
}

\begin{figure*}[!ht]
\centering
\includegraphics[width=10cm]{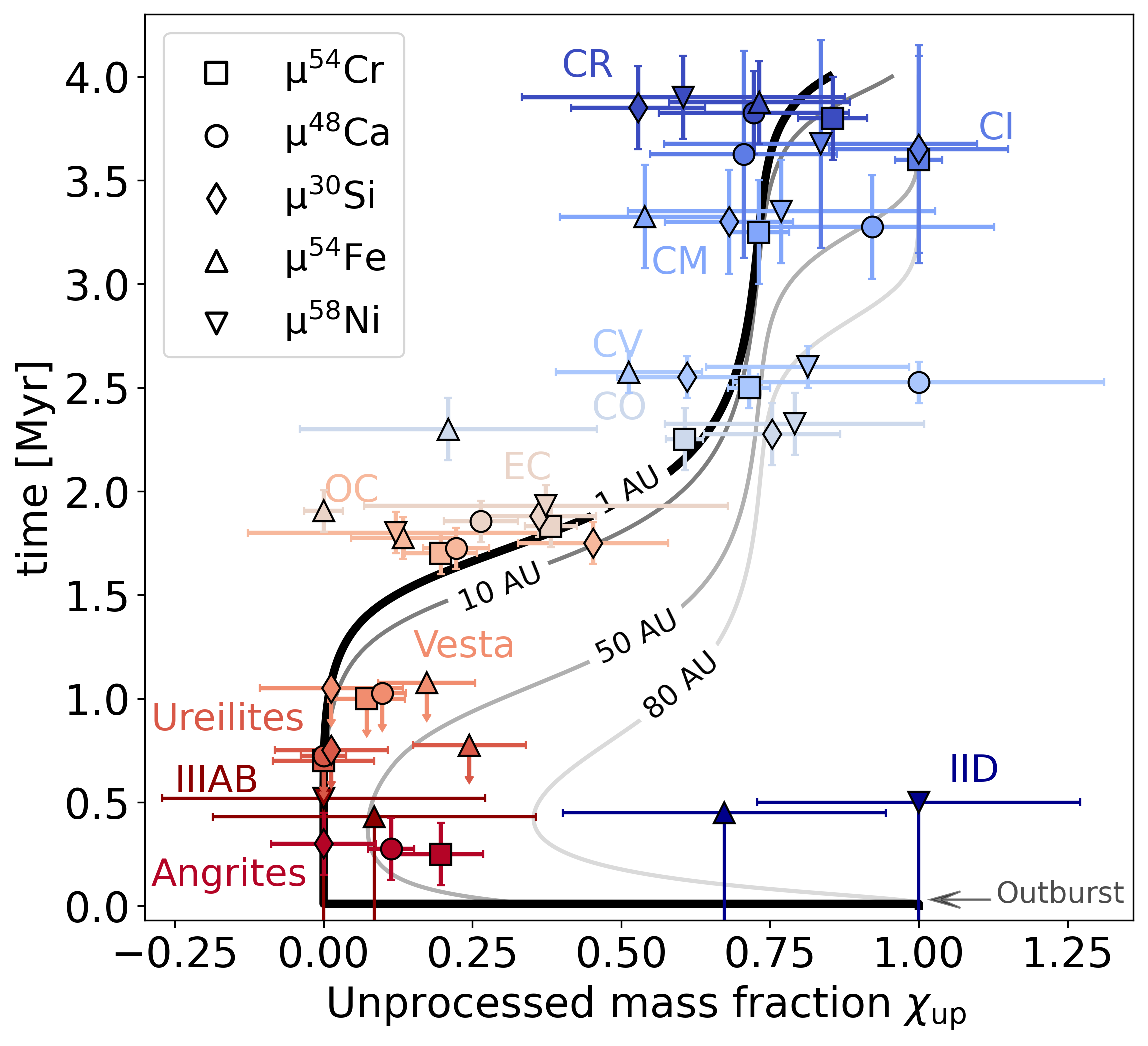}
\caption{
Evolution of the mass fraction of unprocessed pebbles with time at different locations in the disk (contours). The bold black line represents the inner disk composition at 1\,AU. The symbols show normalized nucleosynthetic measurements, for different elements as indicated by the legend. The different meteorite classes are colored from red to blue. The corresponding estimates for their accretion ages are given with vertical error bars (or upper limit arrows) {, slightly displaced for the different elements for visibility}. 
Horizontal error bars are the uncertainties on the normalized nucelosynthetic composition. 
}
\label{fig:pops_all}
\end{figure*}

We consider, for each meteoritic class, the nucleosynthetic composition for a collection of different elements. 
Specifically, we consider here the isotope variations in 
µ\ce{^{54}Cr},
µ\ce{^{30}Si},
µ\ce{^{48}Ca},
µ\ce{^{58}Ni}, and 
µ\ce{^{54}Fe}.
 {
Here the µ-notation reflects the standard normalized isotope ratio expressed in parts per million (see Appendix\,\ref{app:upmassfrac} for definition and used isotope ratio).
\autoref{tab:nucleosyn} gives literature measurements for the different meteorite classes.
}
These isotopes mainly have supernovae (SN) origins, but the precise nature of the SN dust carriers remains unclear.
However, at least for \ce{^{54}Cr} it is understood that SN-produced chromium-rich oxide (chromite) presolar grains contribute to observed nucleosynthetic variations \citep{dauphas2010}.
For Ni, we choose the neutron-poor isotope \ce{^{58}Ni}, which is the main carrier of the isotopic anomaly
 {with a SN origin} \citep{Steele2012,Makhatadze2023}.
{The selected elements are chosen to span a range of condensation temperatures, to include both lithophile (silicate mantle-loving) elements as well as siderophile (iron core-loving) elements, and to sample iron-peak elements.}
The  {chosen} elements are all refractory, in order of increasing 50\%-condensation temperature: 
Cr ($T_{\rm c}= 1291$\,K),
Si ($T_{\rm c} =1314$\,K),
Fe ($T_{\rm c}= 1338$\,K),
Ni ($T_{\rm c}= 1363$\,K),
Ca ($T_{\rm c}= 1535$\,K), 
following \citet{wood2019}. 
We both include lithophile elements (Cr, Si, Ca), as well as siderophile elements (Ni, Fe) \citep{lodders2003}. 
Finally we note that Cr, Fe and Ni are so-called iron-peak elements near the maximum in nuclear binding energy and that Si is special in the sense that is the the most abundant non-volatile element in the Solar System.

{
In this work we will not specifically focus on s-process sensitive elements, like siderophile Mo or lithophile Zr, that are are not major planet-forming elements \citep{Budde2016,burkhardt2021,Spitzer2020}. Their isotopic imprint requires a more complex interpretation due to s-process isotopic variation that is likely carried by AGB-produced SiC stardust grains. However, for completeness, we do present in \autoref{app:Mo} our analysis for the well-studied element Mo, which appears to be consistent with outburst-driven thermal processing, as explored below. 
}

\subsection{A comprehensive analysis}
\label{sec:companalysis}

In order to collectively show the nucleosynthetic anomalies in different elements, we rescale the different nucleosynthetic anomalies such that the unproccesed mass fraction $\chi_{\rm up}=1$ represents the the high µ-value end-member and $\chi_{\rm up}=0$ represents the low µ-value end-member, corresponding to thermally altered material (\autoref{sec:xupsection}).

The choice of the end-members is based on the most extreme values measured in different chondrite and achondrite groups which can be found in \autoref{tab:nucleosyn}.
In principle, other choices for end-member compositions could be made. 
The high µ-value end-member could for example be matched to CAI\slash AOA refractory inclusions. 
{Or, rather, to a hypothetical version with chondritic composition \citep{burkhardt2019}, although the isotopic diversity between CAIs presents a further challenge \citep{Larsen2020}.
However, in practice such a procedure} would only achieve a rescaling of the $\chi_{\rm up}$ parameter.
A more detailed description of this re-scaling approach is given in Appendix\,\ref{app:upmassfrac}. Using this method, we conveniently remain agnostic to the exact nature of the SN-made carriers of the isotopic imprint present in unprocessed material and how exactly these nanoparticles are lost during the thermal processing of particles with icy mantles.

Under the assumption that the low µ-value end-member composition maps to fully thermally processed dust, we find that an outburst model would be consistent {with} the nucleosyntethic evolution seen in Solar System meteorites.
As discussed in \autoref{sec:outburst}, models without outbursts are typically not as efficient in thermally processing a large mass fraction of dust  {and therefore do not match the observed nucleosynthetic evolution (see \autoref{app:extramodels}, \autoref{fig:nom_comp})}. {We therefore here focus on models that include accretion outbursts.} 
{We first consider the evolution of the pebble composition for the model with the strongest outburst (\texttt{Outburst-NoVisc}). Figure \ref{fig:pops_all} shows the evolution of the pebble composition, with time following the vertical y-axis and the unprocessed mass fraction on the horizontal x-axis.
} Different contour lines trace the composition at different locations in the disk, with the thick black line showing the inner-disk composition at 1 AU.
 
Initially the disk is made out of pure pristine unprocessed material. In model \texttt{Outburst-NoVisc}, an outburst after 0.01\,Myr processes the material in the disk out to $70$\,AU. 
This causes the near-horizontal line at the bottom of the plot, illustrating the large-scale thermal processing of disk material. {After the outburst, the viscously expanding gas disk pushes processed material outward, as discussed in \autoref{results}. 
}
Later, after approximately $1.5$\,Myr, the inner disk composition gets slowly taken over by inward-drifting unprocessed pebbles coming from the outer disk ($r\gtrsim$100\,AU), which causes the transition from processed back toward an unprocessed composition after the outburst.
In contrast, the composition of the outer disk ($r\gtrsim$50\,AU) undergoes less thermal processing and looses this signature more efficiently due {to} inward pebble drift.

To compare the nucleosynthetic evolution, the overplotted symbols show the composition of different meteorite classes (indicated with labels and different colors).
Differentiated early-formed bodies (ureilite, angrite, Vesta HEDs) collectively have a  {low µ-value} signature in line with being formed from heavily thermally processed solids present early in the inner disk.
The exception are IID iron meteorites with a  {high µ-value} imprint in \ce{^{58}Ni} and \ce{^{54}Fe} \citep{nanne2019,hopp2022}. 
These could be consistent with the model, if they indeed formed prior to the thermal processing event, which has then to occur before the upper limit on their accretion age \citep[$\leq 0.5$\,Myr,][]{kruijer2017}.  {In fact, iron meteorites come with a wide range of isotopic compositions intermediate between the IIIAB and IID end members shown in \autoref{fig:pops_all} \citep{nanne2019,hopp2022}. For completeness, the composition range of the iron meteorites in \ce{^{58}Ni} and \ce{^{54}Fe} is shown in panels d and e of \autoref{fig:pops2}, in our view illustrating the rapidly evolving inner disk composition during the outburst phase.}

Later formed ordinary {and enstatite} chondrite parent bodies have an isotopic composition in between the low µ-value and  high µ-value end-members, corresponding to an epoch in the disk model where processed and unprocessed pebbles are well mixed in the terrestrial planet forming zone around $1$\,AU. 
Their existence suggests that no strict dichotomy existed between, one the one hand, the carbonaceous chondrites and on the other hand the differentiated, OC {and EC} parent bodies -- that are part of the so-called NC group, as earlier proposed by for example \citet{warren2011,kruijer2020,burkhardt2021}.

Finally, carbonaceous chondrites represent late-formed bodies ($\gtrsim$2\,Myr). In our model, their formation locations are not constrained, as location contours naturally tend to a final unprocessed composition at all locations in the disk. Even CI-like bodies could form in the inner disk, as long as they form late in the evolution of the disk.

\begin{figure*}[!ht]
\centering
\includegraphics[width=16cm]{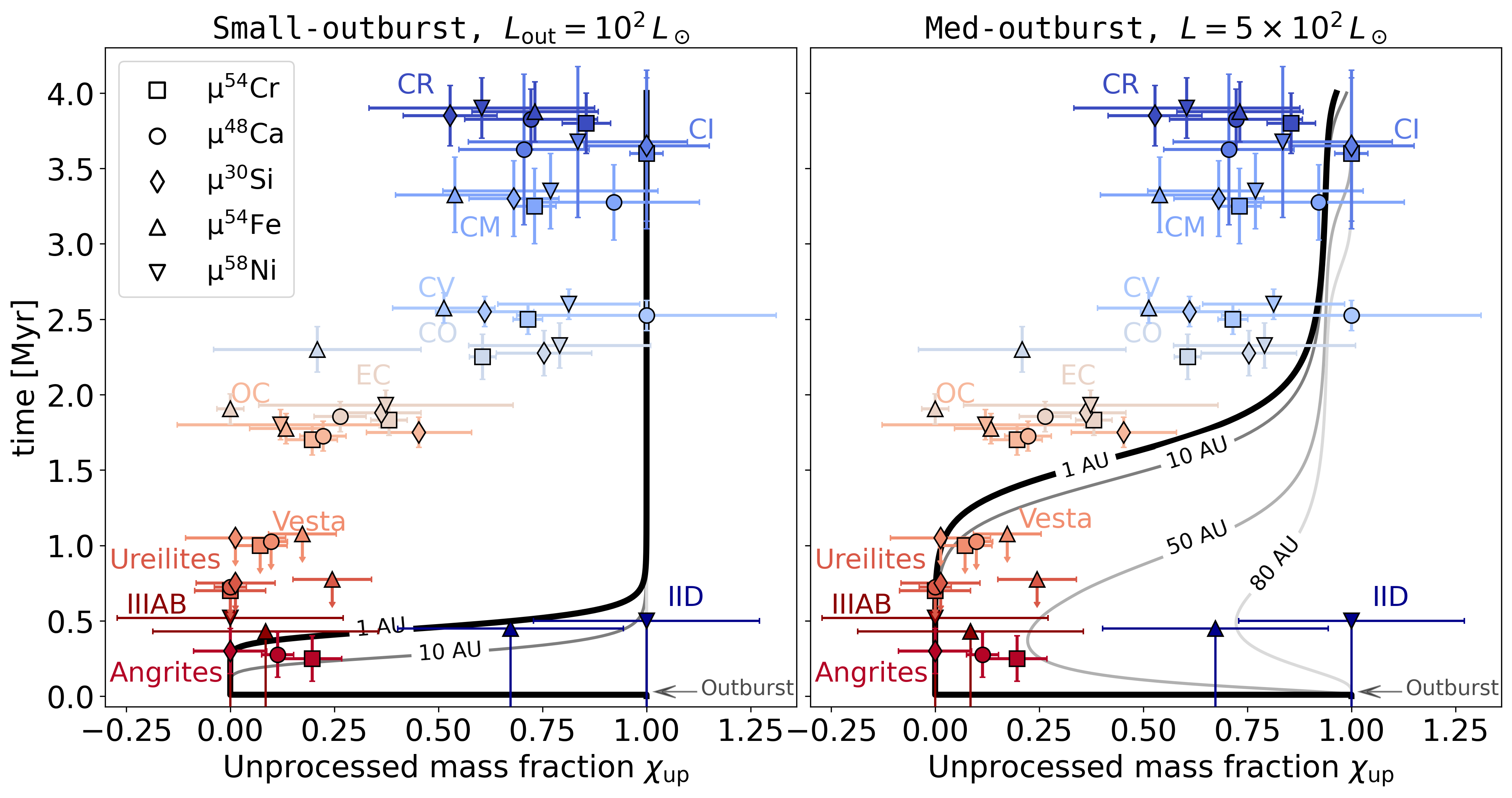}
\caption{{Same as \autoref{fig:pops_all}, but for models with less intense outbursts (\texttt{Small-outburst} and \texttt{Med-outburst}). The outbursts in these cases are not strong enough to push the iceline sufficiently outward. This prevents a long-lived mixing of the processed and unprocessed components, which in our model is needed to match the isotopic differences.}}
\label{fig:low_outburst}
\end{figure*}

The relatively large spread in the nucleosynthetic composition seen in the carbonaceous chondrites may be due to the heterogeneous composition of chondrites composed of different mass fractions of fine-grained matrix, chondrules, and Calcium-aluminium-rich inclusions (CAIs). 
For example, the CC-group shows a large spread in µ\ce{^{48}Ca}, with CV, rather than CI, being the high µ-value end-member. This may be due CI chondrites lacking the presence of CAIs enriched in the refractory species \ce{^{48}Ca} 
\citep[][Appendix \autoref{fig:pops2}, panel a]{schiller2015}. 
Similarly, the CI chondrites are not a high µ-value end-member for µ\ce{^{54}Fe}. Instead CI is highly \ce{^{54}Fe}-poor, possibly due to the lack of chondrules in CI that may have been the dominant metal component carrier 
\citep[][Appendix \autoref{fig:pops2}, panel e]{schiller2020}. 
We also note that the measurement of the isotopic anomalies in Ni show large differences, for example for CI, between different works \citep{nanne2019,Makhatadze2023}.
Other element choices, µ\ce{^{54}Cr}, µ\ce{^{30}Si}, µ\ce{^{58}Ni} do show CI as the expected high µ-value end-member.
{
In particular, CV chondrites contain µ-value rich CI-like matrix that includes chondrules that carry a µ\ce{^{54}Cr} poor signature, similar to NC achondrites, consistent with their precursor formation in the processed inner disk as proposed here \citep{vankooten2021}.
}
Hence, although we argued here that thermal processing of pebbles could be a dominant factor in determining the nucleosynthetic composition of different Solar System objects, it also appears that an additional component is the transport and inclusion of material in the form of chondrules and refractory inclusions \citep{vankooten2021,Hellmann2023}.
Therefore, an interesting avenue for future work would be to extend the model to also include explicitly the contributions of early-formed anomalous CAIs to the final bulk isotope composition.
Their high µ-value nature could be explained if they selectively acquired the sublimated dust fraction during the thermal processing event \citep{trinquier2009,Larsen2020}.

{For completeness, we also briefly discuss models where we consider lower-luminosity outbursts.
As discussed before (\autoref{sec:outburst}), the increase in luminosity in models \texttt{Small-outburst} and \texttt{Med-outburst} is not high enough to efficiently push the icelines outward, which results in less overall thermal processing.
For the $100\, L_\odot$ outburst, material gets processed out to only $\sim$10 AU and quickly gets replenished by the inward drift
of unprocessed pebbles. Therefore, all late-formed differentiated bodies retain an unprocessed composition (\autoref{fig:low_outburst}). By increasing the intensity of the outburst to a luminosity of $500\,L_\odot$, the fraction of processed pebbles increases to $\sim$70\% and results in more outward mixing of processed material, as seen in the right-hand panel of \autoref{fig:low_outburst} This case is able to match the composition of the earliest-formed bodies (angrites, ureilites, HEDs) and some of the late-formed carbonaceous chondrites, but would not be consistent with the composition of the ordinary chondrites and early-formed carbonaceous chondrites (CO and CV).}

To conclude, a strong large-scale thermal processing event, triggered by a stellar outburst, appears to be consistent with the interpretation that the nucleosynthetic isotope differences between meteorites are due the temporal evolution of the composition of the disk, rather than through a fixed spatial separation of different planetesimal reservoirs. 
As opposed to this latter hypothesis, the early inward drift of {low µ-value} material naturally explains the lack of late-formed parent bodies with a {low µ-value} signature (the empty upper left-hand corned of \autoref{fig:pops_all}).
The interpretation of the siderophile elements, \ce{^{58}Ni} and \ce{^{54}Fe}, is more complex, but the  {high µ-value} and  {low µ-value} end-member compositions of iron meteorites may be due their formation, respectively, before or after the last major outburst that triggered a near disk-wide thermal processing event.

\section{Discussion and future work}
\label{sec:discussion}

\subsection{Particle size dependency}
\label{sec:parsize}
 {
In this section we explore the effect of changing the size of the solid particles in the \texttt{Outburst-NoVisc} model.  We consider three scenarios, two with different constant particle sizes ($R_{\rm p}=0.01$\,mm and $1$\,mm) 
and one with varying size between {both} pebble populations {($up$ and $p$)}. 
}

\begin{figure*}[!ht]
   \centering
   \includegraphics[width=18cm]{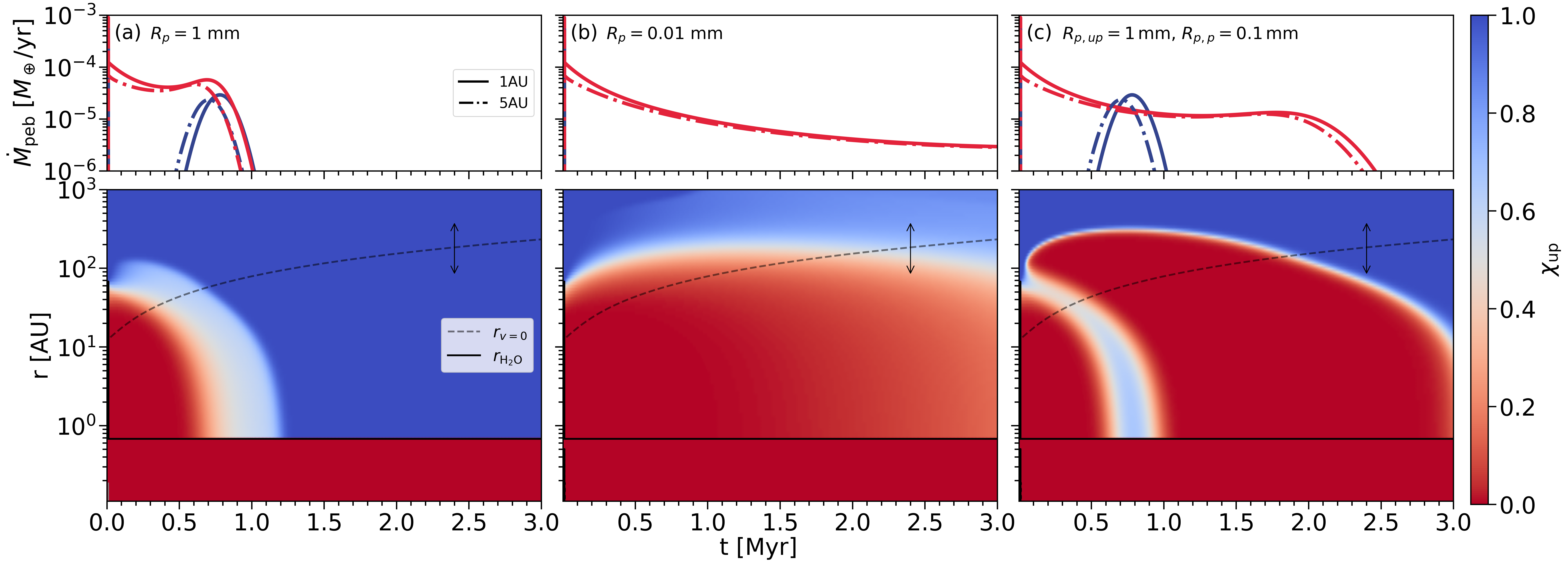}
   \caption{
   Pebble flux evolution for the \texttt{Outburst-NoVisc} model, but using different particle sizes. Panel (a) shows the case with $R_{\rm p}=1$\,mm particles, panel (b) the case with $R_{\rm p}=0.01$\,mm particles, and panel (c) shows a simulation with a varying particle size for each pebble component. Increasing the particle size decreases the lifetime of the pebble disk, due to increased pebble drift rates. If the different pebbles species have different particle radii, large particles in the outer disk can drift inward before smaller processed pebbles are first diffused outward and subsequently later drift inward.
   }
    \label{Conmap3}
\end{figure*}

 {
For the bigger particle case ($R_{\rm p} = 1\,\rm mm$), the pebble component drifts inward on a shorter timescale, causing the disk to run out of material faster ($\lesssim $$1.5\, \rm Myr$), as shown in panel (a) of \autoref{Conmap3}. Such rapid dust depletion would be inconsistent with meteoritic ages and protoplanetary disk ages of several Myr that have been inferred with ALMA \citep{ansdell2016,Appelgren2023}. The opposite happens with smaller particles ($R_{\rm p} = 0.01\,\rm mm$), where the longer drift time makes the pebble disk long lived with pebble fluxes higher than $10^{-6}\, M_\oplus/\rm yr$ after $3$\, Myr, as shown in panel (b) of \autoref{Conmap3}. 
Nevertheless, although both scenarios with different particle sizes have a different temporal evolution compared to the \texttt{Outburst} model with $0.1$\,mm-particles, they still show a similar qualitative evolution in their composition, where the inner disk ($\leq10$\, AU) experiences a mixing of the inner processed component with outer unprocessed material drifting inward. 
}

The final case is one where we consider a disk with different pebble sizes for the different populations. This case is presented as a numerical experiment to illustrate the complex behavior when relaxing the assumption of constant particle size. {Here, we have taken the outer unprocessed particles to have a size of $R_{\rm up,p} = 1\, \rm mm$, while the processed component has a size of $R_{\rm p,p} = 0.1\,\rm mm$. This is done to reflect the inferred primitive mm-sized pebbles that make up pristine comets \citep{blum2017}, while chondrules have similar or smaller sizes. This may, in turn, reflect that when material gets heated and turns from unprocessed to processed, it loses part of its mass and might decrease in size. 
}

 {
In this case, the small $p$-pebbles initially flow outward along with the viscous expansion of the gas disk, while the larger $up$-pebbles already drift inward to the larger drag force on them (panel (c) of \autoref{Conmap3}). This process results in a region where the outward moving $p$-pebbles begin to exceed the surface density of depleted $up$ pebbles. The rapid drift of the larger $up$-pebbles creates an enrichment of unprocessed material in the inner part of the disk ($\lesssim$$10\,\rm AU$). This enhancement is maintained for $\sim$$0.5\, \rm Myr $. However, once this unprocessed component has completely drifted inward and been accreted into the star, the disk starts to follow a similar behavior as before, where the processed component dominating most of the inner disk. This brief early period of $up$ enrichment would allow bodies forming in the inner disk to accrete a fraction of unprocessed material at earlier times compared to the models previously presented.    
}

The above numerical experiments illustrate how our results may depend on a more detailed treatment of the particle size evolution. Therefore, it will be important to explicitly investigate the role of particle coagulation, compaction and fragmentation \citep{Guttler2010,Zsom2010}. Future work will need to go beyond the simplified constant pebble size model used here, which only approximates that, in fact, most mass in solids is carried in pebbles limited in size by fragmentation \citep{brauer2008,birnstiel2010,appelgren2020,Appelgren2023}.
This latter process was believed to be less efficient for icy particles \citep{Gundlach2015}. However, recent laboratory experiments argue fragmentation velocities are not much influenced and remain around $1$\,m$\, \rm s^{-1}$ \citep{Musiolik2016a, Musiolik2016b}. 
In particular, during the outburst, particles undergoing water sublimation may undergo complete destruction down to the monomer level, before recoagulating \citep{Schoonenberg2017,Houge2023}. {However, dust modeling of V883 Ori argues against such aggregate destruction during an outburst \citep{Houge_2023}.}

Collectively, these {coagulation and fragmentation} processes will redistribute, possibly preferentially, isotope carriers along the particle size distribution,
especially when nano-sized particles are trapped below an electrostatic barrier for particle growth
\citep{Akimkin2023}. 
Moreover, the subsequent reduced drift rates of gas-coupled stardust grains, compared to larger pebbles, will further separate these components and leave an imprint in the global disk composition \citep[][]{Steele2012, pignatale2017,hutchison2022}.
Taken together, this is an important avenue for future work.

\subsection{Processes impacting the global pebble flux}
\subsubsection{An initially outward expanding disk}

A fundamental element of the work presented here is that the disk undergoes an initial phase of strong outward  expansion. 
This early expansion appears to be supported by observations of dust radii of class 0/I young disks \citep{maury2019}.
Here we made use of a simplified viscous expansion model  (\autoref{sec:methodsgas}), but determining the exact nature of the outward expansion of dust in disks is of key importance, as it regulates the pebble flux at late times \citep{liu2022}.

{Recent non-ideal magnetohydrodynamic (MHD) simulations illustrate disk where inward mass transport is driven by disk winds also can show outward expansion, due to reconnecting poloidal field lines in the outer disk \citep{Yang2021}.}
{In earlier disk phases,} the {newly-formed} massive and compact disks are close to being gravitational unstable, {which would then drive} the initial viscous relaxation {of the outer disk} \citep{Elbakyan2020}. 
{
Moreover, future work is needed to more carefully model the outbursts themselves as the accretion burst itself triggers global angular momentum redistribution \citep{Cleaver_2023}. 
Finally, disks may expand non-viscously. Even after the collapse of the prestellar core, high angular momentum material, with possibly anomalous composition, may be accreted onto the disk in wide orbits through streamers \citep{kuffmeier2023}. }

\subsubsection{Role of Jupiter}

It is uncertain to which degree Jupiter, and the other giant planets in the Solar System, played a role in reducing the pebble flux to the inner disk. 
Even a moderate reduction of the pebble flux can suppress the formation of super-Earths in the inner disk and instead lead to terrestrial planet formation \citep{lambrechts2019a}. 
Possibly, Jupiter reduced the pebble flux to the inner disk, but it is unlikely this happened early, as it appears the rapid accretion of the envelope of Jupiter was halted by disk dissipation \citep{lambrechts2019b}.
 {Moreover, even} gap-forming planets allow for mass transport of pebble fragments  {through diffusion} across its orbit \citep{pinilla2012,draz2019,stammler2023,Kalyaan2023}. 
It therefore seems unlikely that the NC--CC divide in the iron meteorites can be explained solely by an early-formed Jupiter  {that would permanently seal off the inner disk from solid delivery} \citep{kruijer2017}. 
Instead, we have proposed here that this early separation in \ce{^{58}Ni} and \ce{^{54}Fe} is due to sampling parent bodies formed from pre- and post-outburst dust (\autoref{sec:companalysis}). 
 {This is different from the proposal by \citet{liu2022} who suggested that the parent bodies of the CC-like iron meteorites formed early, but far in the outer disk, and could then subsequently be dynamically delivered to the asteroid belt during giant planet formation.}

Nevertheless, although gap-formation by giant planets is not directly required to explain the NC--CC divide, it very well may have left a clear imprint in the transport of less fragmentation-prone species like chondrules and CAIs {\citep{desch2018,haugbolle2019,Jongejan2023}}.
In this context, it is interesting to speculate that CI chondrites formed devoid of strongly thermally-altered material because the fragmentation-resilient component consisting of chondrules and CAIs did not efficiently pass the pressure bumps outside of the Solar System giant planets. 
 {
Such a scenario has also been suggested to explain the observed late addition of CI-like fine-grained dust rims onto earlier-formed inner-disk NC-like chondrule cores \citep{vankooten2021}.
}

\section{Conclusions}
\label{sec:conclusions}

In this work we investigated the thermal processing of pebbles in the earliest phases of the disk lifetime, following \citet{liu2022}.
We first  {considered} a scenario where disks form compact with very small radii ($<10$\,AU), which allows a large fraction of solids to experience elevated temperatures.
However, in this case pebbles are lost due to drift well within the gas disk lifetime, before the formation of carbonaceous chondrites. This occurs even when assuming high levels of viscous expansion of the outer disk (\autoref{compactresults}). 
Moreover, both observations and simulations argue for a more continuous view of gas delivery to young protoplanetary disks \citep{valdivia2022,Kuznetsova2022,kuffmeier2023}. 
 {We then explored how} stellar outbursts, common in the early stages of disk evolution, could have been the trigger to thermally process a large mass fraction of the solids that were present in the inner regions of a larger protoplanetary disk.
 {Our  model shows how the luminosity increase efficiently pushes the separation radius between processed and unprocessed components outward.
We numerically 
} trace how these early-generated different pebble reservoirs evolve with time, separating those pebbles that underwent thermal processing, through water ice sublimation, from those that did not.
We found that a large-scale thermal processing event, followed by an evolving composition of the pebble flux in the inner disk, is largely consistent with the evolving nucleosynthetic signature of Solar System, in line with \citet{schiller2018,liu2022,onyett2023}.
This view does require that the loss of the icy mantles surrounding primitive dust grains leads the preferential removing of grains with an isotope signature of likely supernova origin.
Exploring this assumption, together with an improved treatment of particle growth, sublimation and fragmentation, should be a focus of further investigations.

Overall, our work argues for a view where disks generically are formed with thermally processed solids in the inner disk and pristine icy pebbles in the outer disk. During the lifetime of the gas disk, this latter component drifts and mixes with the inner component. The subsequent formation of planetesimals and protoplanetary bodies in the inner few AU of disks, encompassing the habitable zone, will then inherit the altering elemental and isotopic composition of their pebble-sized building blocks.

\begin{acknowledgements}
{The authors thank the two referees for their valuable comments. M.J.C.\, acknowledges funding from the NASA grant XRP 80NSSC20K0259.} M.L.\,acknowledges funding from the European Research Council (ERC Starting Grant 101041466-EXODOSS). M.L.\,and M.J.C.\,thank the Gunnar och Gunnel K\"all\'ens minnesfond for supporting a 3-month research visit to Lund Observatory. E.v.K. acknowledges funding from the Danish Villum Young Investigator grant (no.  53024). A.J.\,acknowledges funding from the European Research Foundation (ERC Consolidator Grant 724687-PLANETESYS), the Knut and Alice Wallenberg Foundation (Wallenberg Scholar Grant 2019.0442), the Swedish Research Council (Project Grant 2018-04867), the Danish National Research Foundation (DNRF Chair Grant DNRF159) and the Göran Gustafsson Foundation.
\end{acknowledgements}

\bibliographystyle{aa}
\bibliography{tts_master}

\begin{appendix}

\section{Calculation of the unprocessed mass fraction}
\label{app:upmassfrac}

\begin{figure*}
\centering
\includegraphics[width=18cm]{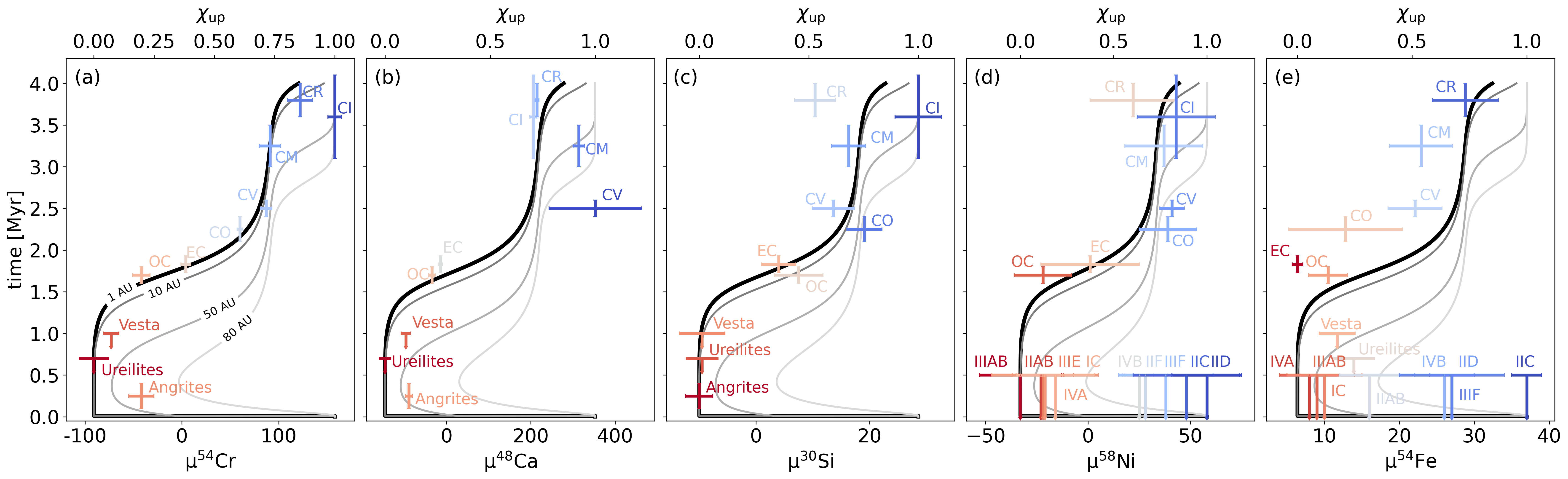}
\caption{Nucleosynthetic measurements for each of the isotopes considered in this work. Each panel represents a different element. The different meteorite classes are red to blue colored, and the horizontal error bars are the uncertainties on the nucelosynthetic composition. The corresponding estimates for their accretion ages are given with vertical error bars (or upper limit arrows). The upper axis represents the normalized values, similar to \autoref{fig:pops_all}. The normalization is discussed in the text. The contours show the evolution of the mass fraction of unprocessed pebbles with time at different locations in the disk. The bold black line represents the inner disk composition at 1AU.}
\label{fig:pops2}
\end{figure*}

{In \autoref{implications} we expressed the nucleosynthetic measurements for different isotopes  in a normalized form by defining the unprocessed mass fraction $\chi_\mathrm{up}$. We give a more detailed description here.}
Isotopic anomalies are typically expressed with either so-called µ notation, defined as follows
\begin{equation}
    \text{µ}^{i}\mathrm{X} = \bigg[\frac{(^{i}\mathrm{X}/^{j}\mathrm{X})_\mathrm{sample}}{(^{i}\mathrm{X}/^{j}\mathrm{X})_\mathrm{standard}}-1\bigg]\times 10^6
\end{equation}

Here, $i$ and $j$ correspond to 54 and 52 in the case of Cr, 48 and 44 for Ca, 58 and 61 for Ni, 30 and 28 for Si, and 54 and 56 for Fe, respectively. All isotope measurements considered in this work are listed in \autoref{tab:nucleosyn} and can be inspected, for each element, in \autoref{fig:pops2}.  {The values for the additional iron meteorites shown in \autoref{fig:pops2}, panels (d) and (e), are taken from \cite{nanne2019} and \cite{hopp2022}, respectively.}

For calculating the unprocessed mass fraction, $\chi_\mathrm{up}$, for an element $X$, we use the following definition
\begin{equation}
    \chi_{\mathrm{up},i} = \frac{\text{µ} X_i -  \text{µ} X_\mathrm{dep}}{ \text{µ} X_\mathrm{enr}- \text{µ} X_\mathrm{dep}}\,,
\end{equation}
where µ$X_\mathrm{dep}$ and µ$X_\mathrm{enr}$ represent the measurement for the  {low µ-value} and  {high µ-value} end-members, respectively.

To select the end-members, we choose the bodies that have the two most extreme values. For example, for µ\ce{^{54}Cr}, the ureilites were chosen as the \ce{^{54}Cr}-poor end-member, while the CI chondrites were taken for the \ce{^{54}Cr}-rich end-member. The end-members of each isotope are indicated in \autoref{tab:nucleosyn}. In the case of µ\ce{^{58}Ni}, both end-members correspond to iron meteorites. For the other isotopes the  {high µ-value} end-member usually corresponds to undifferentiated chondrites like CV and CI for µ\ce{^{48}Ca} and µ\ce{^{30}Si}, respectively. However, the measurement of µ\ce{^{54}Fe} for CI chondrites is highly \ce{^{54}Fe} poor, therefore we take {enstatite chondrite} as the low µ-value end-member.  

{As a final step}, we also propagate the uncertainties from the end-members into the errorbars of \autoref{fig:pops_all}. For this, we simply consider that the error is
\begin{multline}
      \sigma_{\chi_{\mathrm{up},i}}^2 = (\text{µ} X_\mathrm{enr}-\text{µ} X_\mathrm{dep})^{-4}\bigg[\sigma_\mathrm{enr}^2(\text{µ} X_\mathrm{dep}-\text{µ} X_i)^2+\\
    \sigma_\mathrm{dep}^2(\text{µ} X_\mathrm{enr}-\text{µ}X_i)^2+\sigma_i^2(\mu X_\mathrm{enr}-\text{µ} X_\mathrm{dep})^2\bigg] 
\end{multline}

here $\sigma_\mathrm{enr}$, $\sigma_\mathrm{dep}$, are the uncertainties for the end-members.

\section{Unprocessed mass fraction of the non-outburst models}
\label{app:extramodels}

\begin{figure}
\centering
\includegraphics[width=9cm]{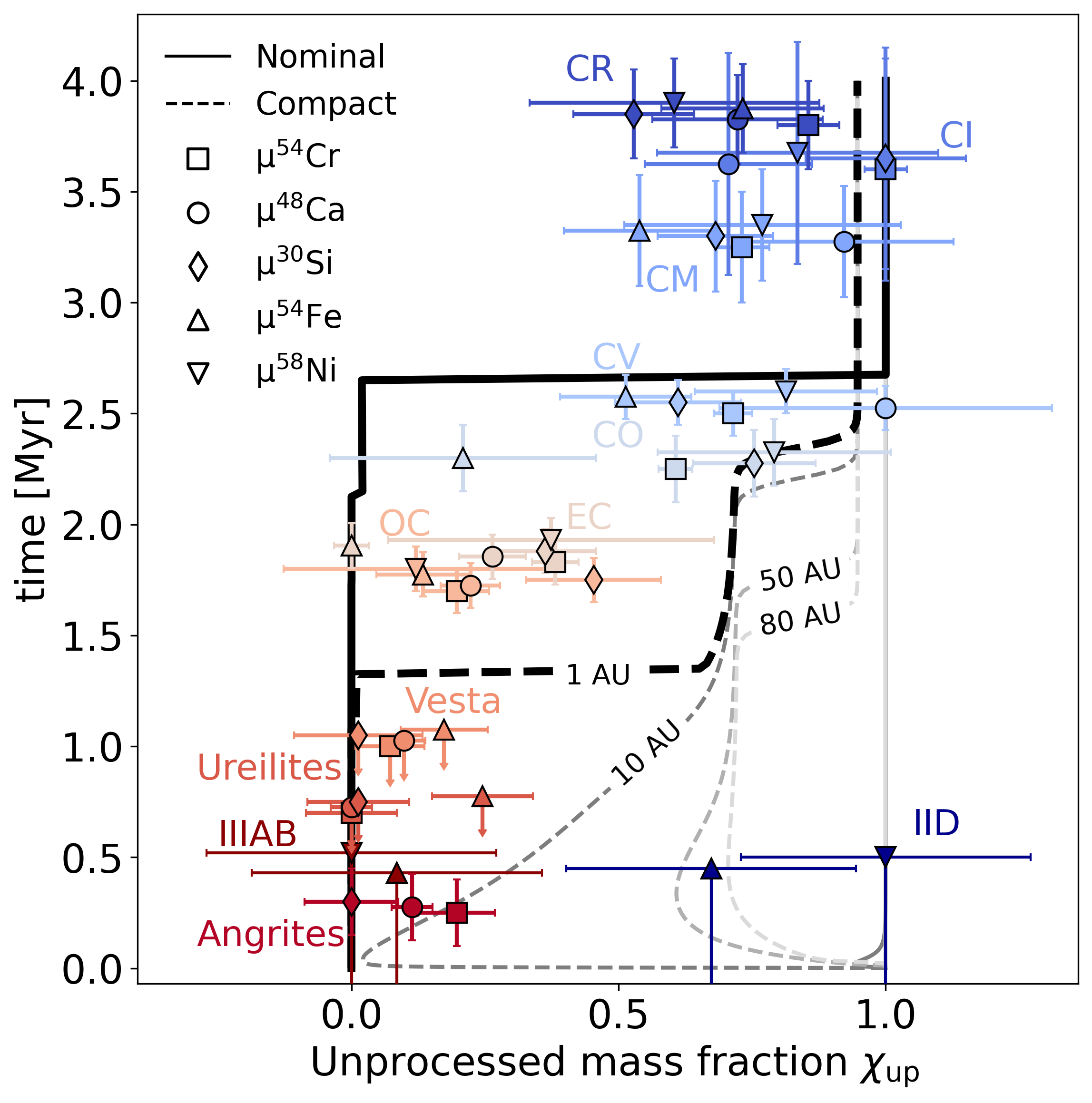}
\caption{ {Evolution of the mass fraction of unprocessed pebbles for the \texttt{Nominal} and \texttt{Compact} models (similar to \autoref{fig:pops_all}). Here, the \texttt{Nominal} model is shown with solid lines (thick black line is 1 AU, the subsequent gray lines show the 10\,AU, 50\,AU and 80\,AU contours), while the dashed lines show the \texttt{Compact}  model. 
}
}
\label{fig:nom_comp}
\end{figure}

{For completeness, we show in this section the evolution of the pebble composition for the \texttt{Nominal} and \texttt{Compact} models in \autoref{fig:nom_comp}. As discussed in section \ref{sec:xupsection}, in the \texttt{Nominal} model a sharp transition from processed to unprocessed material occurs at approximately $2.7$\,Myr at a radius of $1$\,AU shown by the black solid line. This makes it not longer possible to create bodies with a mixed OC-like composition. The \texttt{Compact} models are traced by the dashed line. This case initially show a higher degree of mixing that subsequently gets dominated by the drift of unprocessed material after $\sim$2.5\,Myr, which makes it difficult to match to composition of ordinary chondrites.}

\section{Outburst time}
\label{app:outbursttime}

\begin{figure*}[t!]
\centering
\includegraphics[width=18cm]{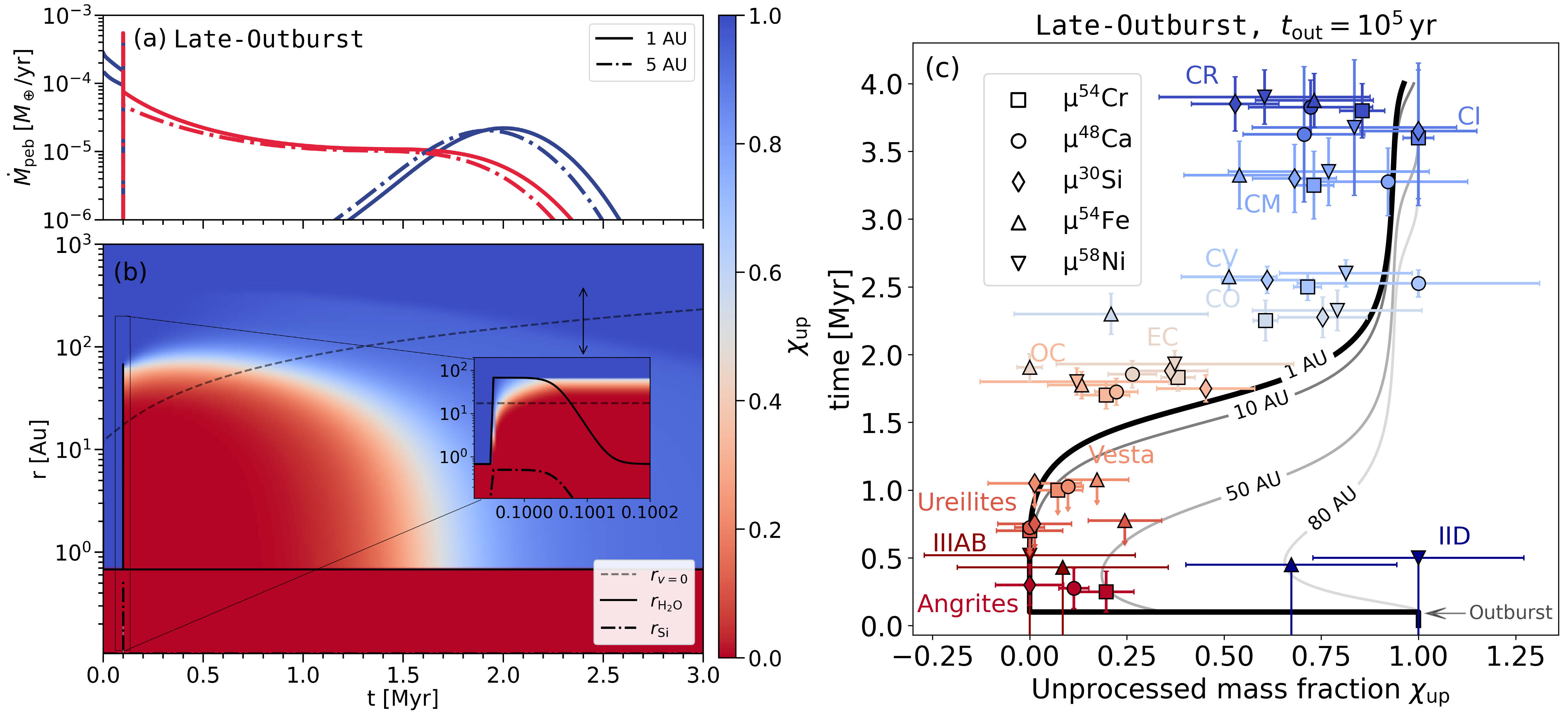}
\caption{{Panels (a) and (b): evolution of the mass fraction of unprocessed pebbles for model \texttt{Late-Outburst}, where the outburst is introduced at $t_{\rm out}=10^5$\,yr (similar to \autoref{Conmap1}). Panel (c): evolution of the mass fraction of $up$ pebbles compared to isotopic compositions of meteorites for the \texttt{Late-Outburst} model (similar to \autoref{fig:pops_all}).}} 
\label{fig:late_outburst}
\end{figure*}

{As discussed before in \autoref{sec:outburst}, the timing, duration and intensity of accretion outbursts change substantially between objects. Outburst are observed to occur typically with the first Myr of disk evolution and last days to centuries \citep{hartmann1996,fischer2023}. Therefore, in this section we briefly explore if a late-inserted outburst can be consistent with the evolution of the nucleosynthetic composition of meteorites.
}

{For this purpose, we explore a new model \texttt{Late-Outburst}, where we use the same parameters as in model \texttt{Outburst-NoVisc}, but modify the age at which we introduce the outburst from $t_{\rm out} = 10^4\, \rm yr$ to $t_{\rm out} = 10^5\, \rm yr$. The time evolution of the pebble composition, along with the different meteorite classes, is shown in \autoref{fig:late_outburst} (panel a and b). In this case, the water iceline that separates the processed and unprocessed reservoirs gets pushed to $\sim$$70\,$AU, similar to model \texttt{Outburst-NoVisc}. However, by the time the outburst is introduced at $10^5\,\rm yr$  a lot of the material has already drifted inward and been accreted into the star.  This means that less material gets processed and that there is a lower degree of mixing, resulting in the composition to be more $up$-like. From \autoref{fig:late_outburst} (panel c), we can see that the disk would still be able to form achondrite-like bodies early on, but the composition after $1.5$\,Myr would be too unprocessed to form bodies with an intermediate composition that is similar to ordinary, CV and CO chondrites. From this we can conclude that outbursts are most efficient in thermally processing material in the compact young disks with which FU Ori outbursts are typically associated \citep{Kospal2023}.
}

\section{Viscosity variation}
\label{app:alphavar}

{The distribution of angular momentum in the disk is determined by the viscous parameter $\alpha$ (\autoref{eq:visc}) in our simulations. 
In the previously presented models, we selected a nominal value of $\alpha=0.01$ that aligns with the observed disk lifetimes and measured gas accretion rates \citep{Appelgren2023}. In this section, we show two additional iterations of the model \texttt{Outburst-NoVisc} with lowered viscosity to highlight the impact of decelerating the viscous evolution of the disk.}

\begin{figure*}[!ht]
   \centering
   \includegraphics[width=15cm]{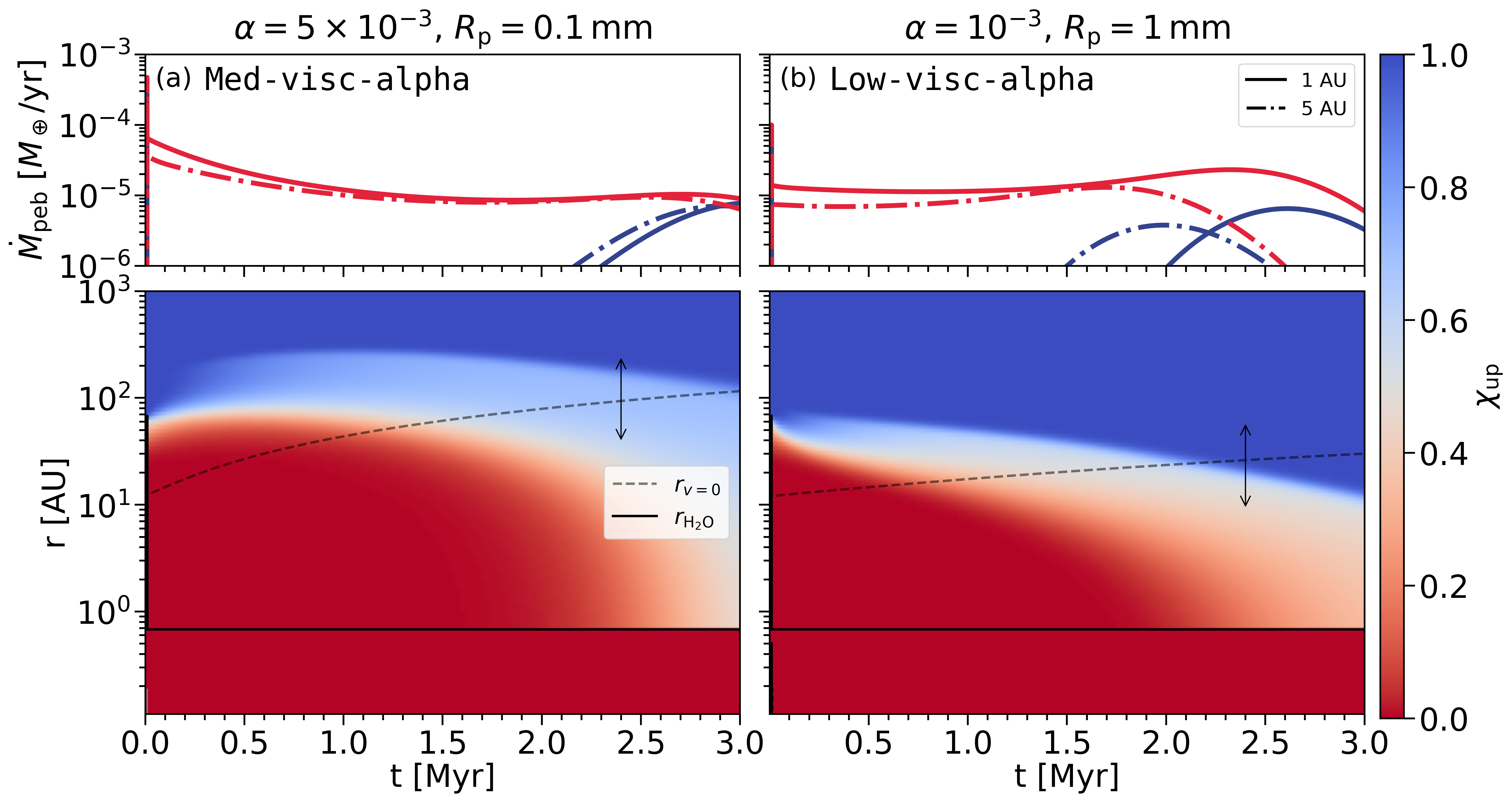}
   \caption{Pebble flux evolution for the \texttt{Outburst-NoVisc} model, but using viscosities of (a) $\alpha=0.005$ and (b) $\alpha=0.001$.}
    \label{fig:alphas}
\end{figure*}

{{We first run the model \texttt{Med-visc-alpha} with $\alpha=0.005$. The evolution of the pebble flux in the disk is shown in panel (a) of \autoref{fig:alphas}. As the viscosity in the disk is decreased, the disk evolves on a slower timescale. Therefore, unprocessed material only becomes dominant inside 5\,AU after 3\,Myr. This is because the lower disk viscosity results in a higher gas surface density at late times which leads to slower pebble drift (for particles with fixed size, see \autoref{eq:drift}). Because of this effect, we run model \texttt{Low-visc-alpha}, where we decrease the viscosity parameter to $\alpha=0.001$, with particle with a larger size of $R_{\rm p}=1\,\rm mm$, promoting drift (\autoref{sec:parsize}). In this way the disk starts to be depleted of pebbles within a few million years, more consistent with observed disk lifetimes. 
Nevertheless, as shown in in panel (b) of \autoref{fig:alphas}, drift remains relatively inefficient for this particle size choice and the inner disk retains a processed composition within 2.5\,Myr. 
The more rapid decrease of the pebble flux at 5\,AU, compared at 1\,AU, illustrates the depletion of a smaller reservoir of outer-disk solids, due a lower degree of initial outward-drifting pebbles during the viscous expansion phase of the disk and subsequent drift.
In summary, the evolution of the disk composition is only moderately sensitive to the choice of the viscosity for the initial disk expansion, but the effective Stokes number of the inward-drifting pebbles regulates the delivery of unprocessed material to the inner disk. 

}}

\begin{figure*}[!ht]
   \centering
   \includegraphics[width=15cm]{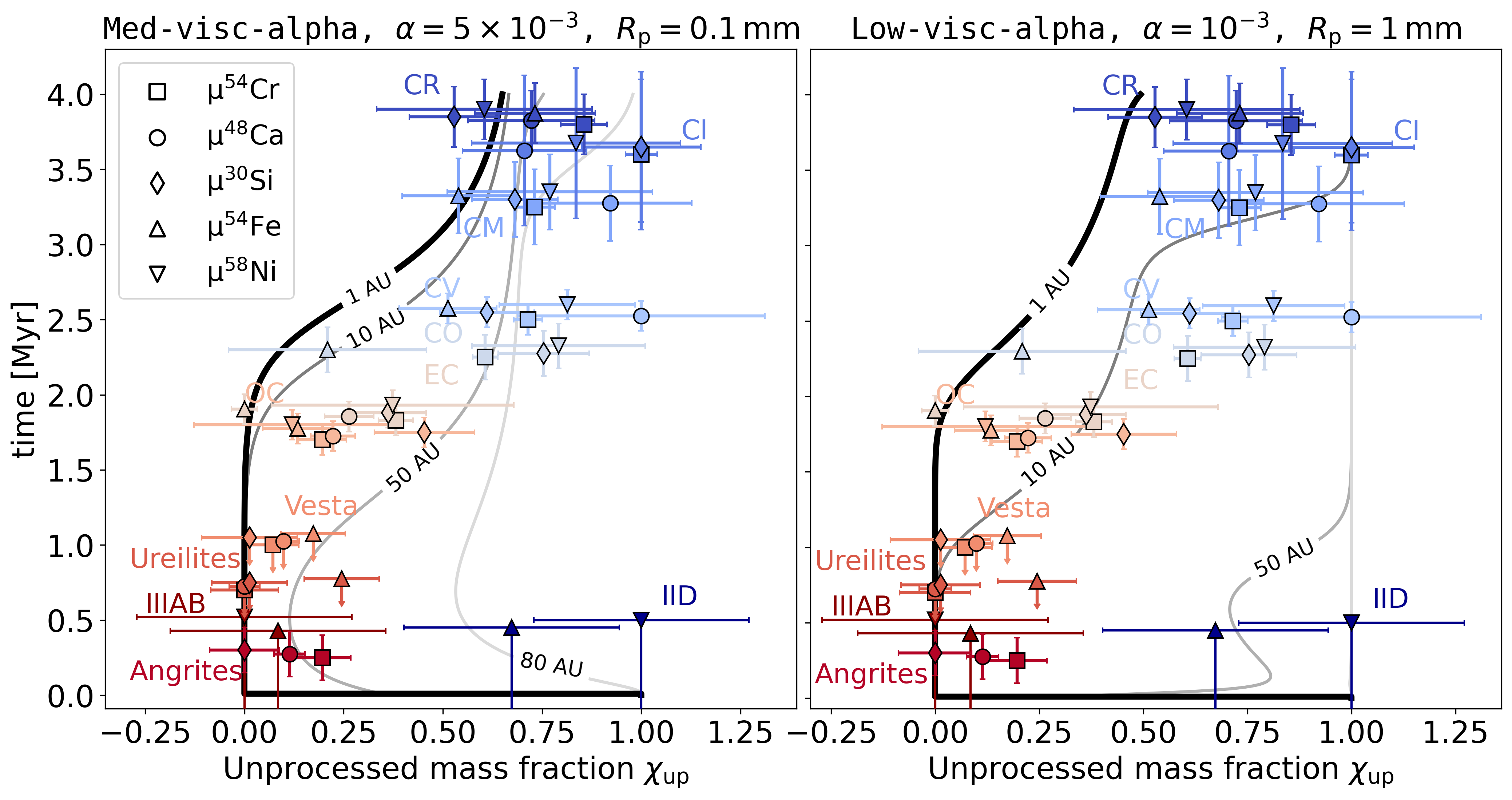}
   \caption{Unprocessed mass fraction evolution for the \texttt{Outburst-NoVisc} model, but using viscosities of (a) $\alpha=0.005$ and (b) $\alpha=0.001$, similar to \autoref{fig:pops_all}}
    \label{fig:alpha_up}
\end{figure*}

{{The slower disk evolution can also be seen reflected in the modeled meteoritic compositions in \autoref{fig:alpha_up}. The model \texttt{Med-visc-alpha} (left panel) is able to match the composition of early-formed, differentiated bodies, but even after $\sim$1.5\,Myr it maintains a high amount of processed material, inconsistent with the isotopic composition of several carbonaceous chondrites. The evolution of the composition of model \texttt{Low-visc-alpha} is slightly more extreme (\autoref{fig:alpha_up}, right panel). Inside $10$\,AU the disk remains dominated by mostly processed pebbles within $2$\,Myr after the outburst. This mainly indicates that the pebble drift has to be more efficient to explain the composition of the carbonaceous chondrites.}}

\clearpage
\section{Molybdenum as an
example of an element with s-process variability
}
\label{app:Mo}

\begin{figure}[!ht]
   \centering
   \includegraphics[width=9cm]{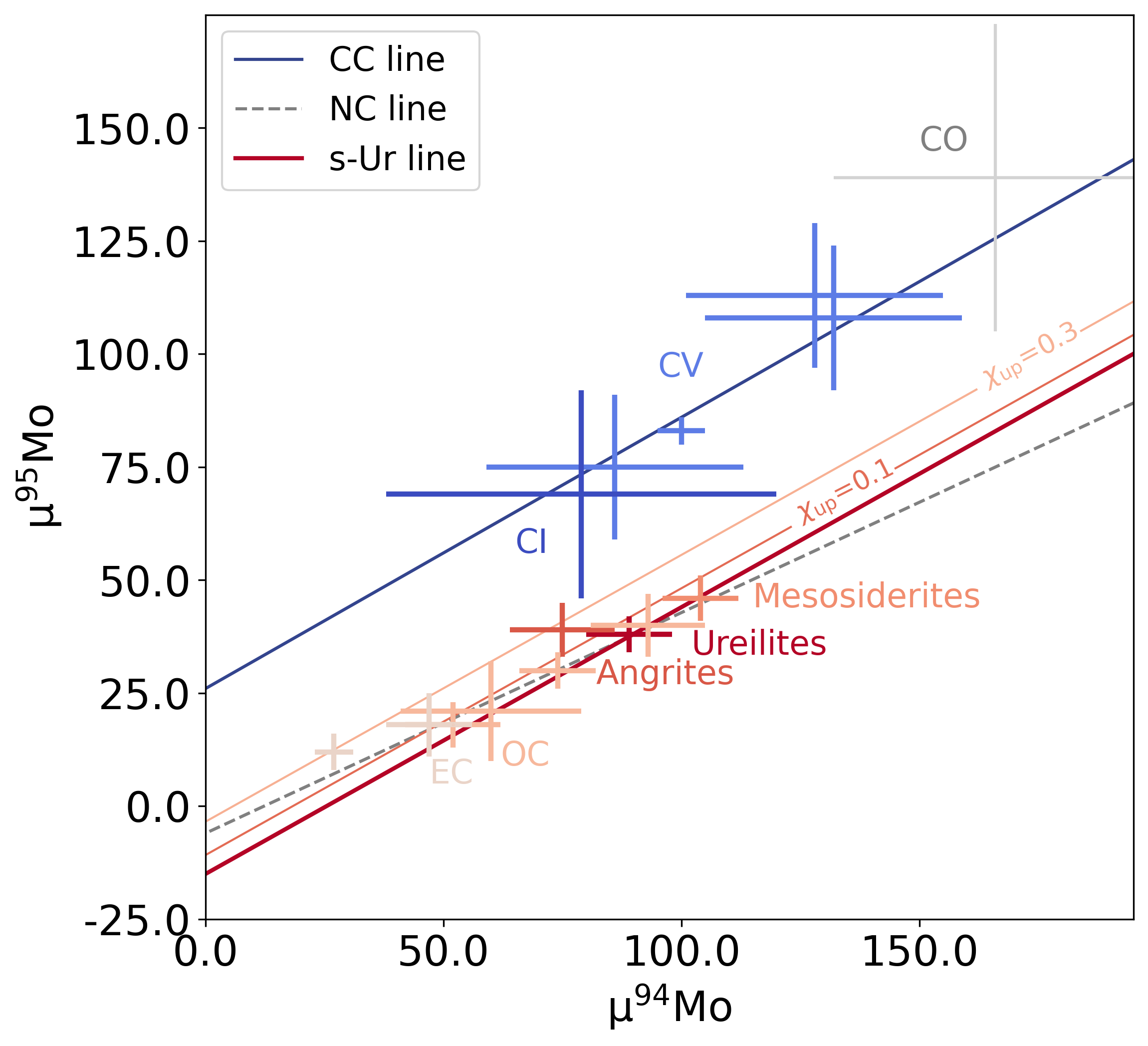}
   \caption{
Mo isotope compositions of the CC and NC meteorite groups in a µ\ce{^{94}Mo} versus µ\ce{^{95}Mo} diagram (using same color scheme as in \autoref{fig:pops_all}). 
The blue CC-line fit is taken from \citet{Budde2016}. 
NC meteorites plot close to an s-process mixing line through Ureilite (thick red s-Ur line). Thin contours give the unprocessed mass fraction encompassing the OC and EC chondrites. The gray dashed line represents the line fit through the NC group obtained by \citet{Yokoyama_2019}.
\emph{Literature data sources}: 
CI \citep{Burkhardt2011},
CO \citep{burkhardt2014},
CV, OC, EC \citep{Yokoyama_2019},
Ureilites, Angrites, Mesosiderites \citep{budde2018}. The latter may represent Vesta-like material \citep{Haba_2019}. 
}
\label{fig:Mo}
\end{figure}

{
Our work focuses on the isotopic variability between different meteorite classes of r-process sensitive elements, likely linked to supernova grains (\autoref{subsec:introclass}). 
In contrast, the different isotopic signatures of the so-called s-process elements have been considered to be more complex, likely due to Silicon Carbide (SiC) grains driving s-process variability, and have been interpreted in multiple different ways \citep{Budde2016,Yokoyama_2019,Stephan2021}. 
Here, we will focus on the well-studied element Molybdenum (Mo) and show that its isotopic signature is also consistent with our interpretation of outburst-driven thermal processing. 
}

{
Molybdenum is a moderately siderophile element that shows both r- and s-process driven variation in bulk meteorites. 
Therefore, a simple analysis with µ-variability in a single isotope, as for example done in \autoref{fig:pops_all}, is no longer instructive. 
Instead, we focus here on a µ\ce{^{94}Mo} versus µ\ce{^{95}Mo} diagram , a so-called three-isotope plot (\autoref{fig:Mo}).
}

{
The different carbonaceous chondrites classes are approximately located on a single line in \autoref{fig:Mo}. 
The blue CC-line fit from \citet{Budde2016} also includes, not shown here, CC achondrites \citep{kruijer2017} and carbonaceous chondrites with strong s-process depletion (high µ values), like Ryugu \citep{Nakanishi2023}, which plot outside the axis range used here.
This CC-line also includes the primitive CI chondrites, marked with a bold blue cross \citep[][although \citet{Dauphas2002b} reports an off-set value]{Burkhardt2011}. 
The substantial s-process variation along the CC-line reflects the heterogeneous distribution of s-process carriers. This has been interpreted as being driven by embedded SiC grains that show Mo isotope variability with the same slope as the CC-line. 
The CC-slope used here has a linear slope of 
$s_{\rm CC}=0.60$ \citep{Budde2016,Yokoyama_2019}. 
SiC grains have a reported s-process slope of $s_{\rm SiC} = 0.59 \pm 0.02$ \citep{Nicolussi1998,Dauphas2004}. 
Alternatively, variability along the blue line may be driven by chondrules carrying a positive anomaly in both  µ\ce{^{94}Mo} and µ\ce{^{95}Mo} compared to matrix, possibly driven by metal-silicate fractionation during chondrule formation \citep{Budde2016}. 
}

{
For the NC group, the isotopic distribution in a µ\ce{^{94}Mo} versus µ\ce{^{95}Mo} diagram is clearly distinct and offset from the CC group (\autoref{fig:Mo}). 
However, it is not readily apparent if the NC distribution is located on a single line with a SiC-like slope, or even on a single mixing line at all. 
Previous works have generally argued for an NC-line with a less steep slope compared to SiC \citep[however see also][]{Stephan2021}. \citet{Budde2016} reported a slope of 
$s_{\rm NC}=0.47 \pm 0.14$, similar to 
$s_{\rm NC}=0.488 \pm 0.085$ that we use here \citep[][gray dashed line]{Yokoyama_2019}. 
Recently, \citet{Spitzer2020} found $s_{\rm NC}=0.528 \pm 0.045$, shallower than the earlier estimate by \citet{budde2018} of $s_{\rm NC}=0.595 \pm 0.011$. 
This shallow slope is driven by the s-process enriched OC end EC samples that lean away from the s-Ur line in the lower left hand corner of \autoref{fig:Mo}.
}

{
Following our outburst model of \autoref{implications}, the CI chondrites (bold blue cross in \autoref{fig:Mo}) correspond to an unprocessed component ( $\chi_{\rm up}=1$). The NC achondrite Ureilite would represent a fully processed endmember ($\chi_{\rm up}=0$).
Therefore, the red s-Ur line with a SiC slope represents an s-process mixing line onto which fully processed solids with SiC-carried heterogeneity would then plot.
We also show parallel curves that represent mixtures between the s-Ur line and the CI chondrite endmember, corresponding to partly processed compositions ($\chi_{\rm up}=0.1$ and $\chi_{\rm up}=0.3$). The ordinary chondrites, in our model a mixture of outburst-processed pebbles and inwards-drifted pristine outer-disk pebbles, fall in this range, 
as well as the enstatite chondrites. 
In summary, the Mo isotope data can be interpreted to be consistent with the progressive loss of isotopically distinct Mo from ice-embedded dust: from unprocessed CI to ordinary chondrite and finally fully-processed Ureilites, as also argued in the main text. 
This interpretation is different from, for example \citet{Budde2016}, that argued for an \emph{addition} of r-process Mo to the CC reservoir. 
That being said, we acknowledge the substantial errorbars on Mo isotopic compositions and alternative explanations cannot be excluded. Future work is needed to further explore how elements with strong s-element variability fit with the thermal processing model.
}

\end{appendix}
\end{document}

%% file: table.tex
    \begin{threeparttable}
    \caption{Ages and isotopic measurements for the meteoritic bodies studied}        
    \label{tab:nucleosyn}
    \begin{tabular}{c c c c c c c c c c c c c} 
    \hline\hline
        ~ & Acc.\, age [Myr] & Ref & µ$^{54}$Cr & Ref & µ$^{48}$Ca & Ref & µ$^{30}$Si & Ref & µ$^{58}$Ni & Ref & µ$^{54}$Fe & Ref \\ \hline
        
        Ureilites & <0.7 & [1] & -90$\pm$15$^\mathrm{a}$ & [9] & -146.2$\pm$13.6$^\mathrm{a}$ & [17] & -9.5$\pm$2.8 & [18] & ~ & ~ & 13.9$\pm$2.8 & [22] \\ 
        
        Angrite & 0.25$\pm$0.25 & [2] & -40$\pm$13 & [10] & -89.5$\pm$8.3 & [17] & -10$\pm$2.4$^\mathrm{a}$ & [18] & ~ & ~ & ~ & ~ \\ 
        
        Vesta/HED & <1 & [3] & -73$\pm$8 & [11] & -97$\pm$10.9 & [17] & -9.5$\pm$4 & [18] & ~ & ~ & 11.7$\pm$2.4 & [22] \\ 
        
        OC & 1.6-1.8 & [4] & -42$\pm$9 & [11] & -34.9$\pm$7.7 & [17] & ~ & ~ & -22$\pm$14 & [21] & 10.5$\pm$2.6 & [22] \\

        EC & 1.83$\pm$0.1 & [5] & 4$\pm$5 & [12]& -14.5 $\pm$4.2 & [18] & 4$\pm$3& [12] & 1$\pm$24 & [21] & 6.4$\pm$0.7 & [22] \\
        
        CO & 2.1-2.4 & [4] & 60$\pm$3 & [11,13] & ~ & ~ & 19.1$\pm$3.1 & [18] & 39$\pm$14 & [21] & 12.8$\pm$7.6 & [22] \\ 
        
        CV & 2.4-2.6 & [4] & 87$\pm$6 & [14] & 353$\pm$110$^\mathrm{b}$ & [19,20] & 13.6$\pm$3.7 & [18] & 41$\pm$6 & [21] & 22.1$\pm$3.6 & [22] \\ 
        
        CM & 3-3.5 & [6] & 91$\pm$11 & [15] & 314$\pm$14 & [17] & 16.3$\pm$3 & [18] & 37$\pm$19 & [21] & 22.9+4.2 & [22] \\ 
        
        CR & 3.6-4 & [7] & 122$\pm$13 & [16] & 214.55$\pm$4.6 & [17] & 10.4$\pm$3.6 & [18] & 22$\pm$21 & [21] & 28.8$\pm$4.4$^\mathrm{b}$ & [22] \\ 
        
        CI & 3.6$\pm$0.5 & [5] & 158$\pm$7$^\mathrm{b}$ & [17] & 206.1$\pm$8.5 & [13] & 28.6$\pm$4.1$^\mathrm{b}$ & [18] & 43$\pm$19 & [21] & -2$\pm$2.7 & [22] \\ 
        
        IIIAB & $<$0.5 & [8] & ~ & ~ & ~ & ~ & ~ & ~ & -33$\pm$20$^\mathrm{a}$ & [21] & 9$\pm$5$^\mathrm{a}$ & [23] \\ 
        
        IID & $<$0.5 & [8] & ~ & ~ & ~ & ~ & ~ & ~ & 58$\pm$17$^\mathrm{b}$ & [21] & 27$\pm$3 & [23] \\\hline







        
        
    \end{tabular}
    \begin{tablenotes}
    \item
    $^\mathrm{a}$ Processed end-member\\
    $^\mathrm{b}$ Unprocessed end-member\\
   $[1]$ \citet{vankooten2017}, [2] \citet{schiller2015}, [3] \citet{schiller2017}, [4] \citet{doyle2015}, 
   [5] \citet{sugiura2014}, 
   [6] \citet{fujiya2012}, 
   [7] \citet{budde2018}, 
   [8] \citet{kruijer2017}, 
   [9] \citet{zhu2020}, 
   [10] \citet{zhu2019}, 
   [11] \citet{trinquier2007}, 
   [12] \citet{Dauphas2024}, 
   [13] \citet{trinquier2006}, 
   [14] \citet{zhu2023} and references therein, 
   [15] \citet{vankooten2020}, 
   [16] \citet{vankooten2016}, 
   [17] \citet{schiller2018}, 
   [18] \citet{onyett2023},
   [19] \citet{moynier2010}, 
   [20] \citet{dauphas2014}, 
   [21] \citet{nanne2019}, 
   [22] \citet{schiller2020}, 
   [23] \citet{hopp2022}.
  \end{tablenotes}
  \end{threeparttable}